\documentclass[lettersize,journal]{IEEEtran}

\usepackage{amsmath,amsfonts,amssymb}
\usepackage{algorithmic}
\usepackage{algorithm}
\usepackage{array}
\usepackage[caption=false,font=normalsize,labelfont=sf,textfont=sf]{subfig}
\usepackage{textcomp}
\usepackage{stfloats}
\usepackage{url}
\usepackage{verbatim}
\usepackage{graphicx}
\usepackage{cite}
\usepackage{bm}              
\usepackage{booktabs}        
\usepackage{multirow}        
\usepackage{placeins}        
\usepackage{tikz}            
\usetikzlibrary{arrows.meta,positioning,calc,fit,backgrounds,
  decorations.pathreplacing}
\usepackage{mathtools}       
\usepackage{enumitem}        
\usepackage{amsthm}          
\usepackage{balance}         
\usepackage[colorlinks=true,citecolor=blue,linkcolor=blue,urlcolor=blue]{hyperref}

\theoremstyle{plain}

\theoremstyle{definition}

\graphicspath{{FIGS/}{section/figures/}}

\begin{document}
\bstctlcite{IEEEexample:BSTcontrol}
\setcounter{page}{1}
\twocolumn



\setcounter{page}{1}
\twocolumn

\title{Mixed-Size Placement Initialization via
  Stiefel-Manifold Sequential Subspace Optimization}

\author{Pengwen Chen,
  Chung-Kuan Cheng,~\IEEEmembership{Fellow,~IEEE},
  Shao-Yun Fang,~\IEEEmembership{Senior Member,~IEEE},
  Chester Holtz,~\IEEEmembership{Student Member,~IEEE},
  Sheng-Tan Huang,~\IEEEmembership{Student Member,~IEEE},
  Susmita Sur-Kolay,~\IEEEmembership{Senior Member,~IEEE},
  Yucheng Wang,~\IEEEmembership{Student Member,~IEEE},
  and~Ying~Yuan$^{\ast}$,~\IEEEmembership{Student Member,~IEEE}%
\thanks{$^{\ast}$Corresponding author: Ying Yuan
  (e-mail: yiy090@ucsd.edu).
}

\thanks{Pengwen Chen is with the Department of
  Mathematics, National Tsing Hua University, Hsinchu, Taiwan
  (e-mail: pengwen@math.nthu.edu.tw).%
}

\thanks{Chung-Kuan Cheng, Susmita
  Sur-Kolay, Yucheng Wang, and Ying Yuan are with
  the Department of Computer Science and Engineering,
  University of California, San Diego, La Jolla, CA,
  USA
  (e-mail: \{ckcheng, ssurkolay, yuw132, yiy090\}@ucsd.edu).%
}

\thanks{Chester Holtz is with
  Halıcıoğlu Data Science Institute,
  University of California, San Diego, La Jolla, CA,
  USA
  (e-mail: chholtz@ucsd.edu).%
}

\thanks{Shao-Yun Fang and Sheng-Tan Huang are with
  the Department of Electrical Engineering, National 
  Taiwan University of Science and Technology
  (e-mail: \{syfang,  d11307009\}@mail.ntust.edu.tw).%
}
\thanks{This work has been submitted to the IEEE
for possible publication. Copyright may be transferred
without notice, after which this version may no longer
be accessible.}
}


\maketitle


\begin{abstract}
Mixed-size global placement simultaneously optimizes the locations of large macros and millions of standard cells in modern very-large-scale integration (VLSI) physical design.
State-of-the-art analytical placers achieve strong scalability, but their nonlinear objectives and density handling make them sensitive to the initial placement.
This paper presents a geometry-driven initialization framework that formulates placement initialization as a Quadratically Constrained Quadratic Program (QCQP) with first-order centroid and second-order spreading constraints.
Explicit mixed-size spreading targets are derived from the die geometry and object sizes, to obtain stable constraint values even under extreme size variation.
A utilization-aware rescaling of the constraint matrix prevents pathological spreading when fixed objects occupy significant die area.
The QCQP is then transformed into a Stiefel-manifold optimization through metric equalization and centering projection, making the underlying geometric structure explicit.
The resulting Stiefel-constrained quadratic objective admits efficient large-scale optimization via a Riemannian sequential subspace method (SSM). 
 An angle-sorted cycle sparsification for multi-pin nets periodically keeps the Laplacian approximation sparse yet geometry-consistent.
As a drop-in initializer for existing analytical placers, our method yields improvements in weighted half-perimeter wirelength (wHPWL): 13.8\% average under RePlAce on the four
TILOS benchmarks supported by its flow, 3.5\% under
DREAMPlace on all 20 benchmarks, and 4.8\% under
AutoDMP on 19 benchmarks with converged baselines.
Post-route evaluation shows that these pre-route improvements carry through to routed designs: routed wirelength decreases by 1.1--2.3\% on all three benchmarks. Standard-cell area and power also decrease on all three. Design-rule-check (DRC) violations decrease on two of the three.
\end{abstract}



\section{Introduction}
\label{sec:intro}

Placement is a central step in very-large-scale integration (VLSI) physical design: given a circuit netlist and a fixed die region, placement assigns locations to circuit components to enable subsequent routing and timing closure.
Modern benchmarks and industrial designs are increasingly \emph{mixed-size}, containing millions of standard cells together with tens to hundreds of macros (e.g., SRAMs, analog blocks, IPs) whose sizes may differ by orders of magnitude.
This disparity in size, coupled with geometric feasibility, density control, and wirelength optimization, exacerbates numerical scaling issues in large-scale optimization~\cite{MarkovHuKim15_ProcIEEE_PlacementSurvey}.

Recent \emph{analytical} placers formulate global placement as a large-scale continuous optimization that balances smooth wirelength approximations with density control and design constraints.
Representative engines include ePlace~\cite{LuCCSHTC15_ePlace} and RePlAce~\cite{ChengKKW19_RePlAce} for electrostatics-based density formulations, ePlace-MS~\cite{LuZhuangChenEtAl15_TCAD_ePlaceMS} for explicit mixed-size handling, DREAMPlace~\cite{LinDLRKP19_DREAMPlace} for GPU-accelerated placement, and AutoDMP~\cite{AutoDMP23_ISPD} for automated parameter tuning in concurrent macro and standard-cell placement.

Although most of these engines scale to very large instances, the overall objective remains nonconvex and sensitive to modeling choices, numerical conditioning, and starting points~\cite{MarkovHuKim15_ProcIEEE_PlacementSurvey, YanViswanathanChu09_DAC_MixedSizeComplexities}.
Prior work explicitly notes that \emph{variable initialization plays a critical role}, especially for quality-focused placement~\cite{ChenCCHLW23_ISPD}.
Learning-assisted flows also report substantial sensitivity to random seeding and parameter settings~\cite{AutoDMP23_ISPD}.
A high-quality initial placement can place the optimizer in a favorable basin of attraction, and under the fixed iteration budgets used by practical placers, this can improve the final placement quality without changing the downstream flow.

Many existing initialization methods for placement can be grouped into two categories.
\emph{Graph-spectral} methods exploit Laplacian-based approximations derived from the circuit netlist, including the constrained spectral optimization via Quadratically Constrained Quadratic Programs (QCQP) introduced by Chen et al.~\cite{ChenCCHLW23_ISPD} and the spectral filtering approach of Liu et al.~\cite{LiuZhouWangEtAl24_GiFt}.
\emph{Learning-based} methods use reinforcement learning or graph neural networks for macro-level layout planning~\cite{Mirhoseini21_Nature_GraphPlacement,Liu22_TODAES_GraphPlanner}.
Our work belongs to the first category and targets a specific gap in constrained spectral initialization for mixed-size circuits.

We investigate whether a structure-aware and constraint-preserving initializer can improve half-perimeter wirelength (HPWL) outcomes over common heuristic initializations in large-scale mixed-size placement.
We build on QCQP formulations, but target stable mixed-size use in the presence of fixed objects and widely varying macro sizes.
Our key observation is that QCQP equality constraints used in area-aware initialization already define a rich \emph{geometric} structure: after metric equalization, the quadratic spreading equalities become an orthonormality condition on a Stiefel manifold \cite{ChenCH24_StiefelSSM_arXiv}.
Once this structure is made explicit, one can apply Riemannian optimization tools and sequential subspace methods (SSMs) for scalable optimization on the constraint set~\cite{AbsilMahonySepulchre08, ChenCH24_StiefelSSM_arXiv}.

Fig.~\ref{fig:method_pipeline} gives a high-level overview.
Starting from a mixed-size netlist, we build a sparse Laplacian approximation and a QCQP.
The QCQP's first-order constraints fix the area-weighted centroid. Its second-order constraints enforce area-weighted spreading, preventing degenerate solutions such as collapse to a single point.
We then derive explicit mixed-size spreading constants, construct a well-conditioned constraint matrix, and transform the QCQP into a Stiefel-constrained optimization via metric equalization and centering projection.
We solve the resulting Stiefel problem by SSM with Riemannian second-order steps, coupled with periodic angle-sorted cycle sparsification for multi-pin nets.

\input{section/figures/fig_pipeline}

Section~\ref{sec:motivation} details the
specific challenges and gaps that motivate each
contribution.
Our main contributions are:
\begin{itemize}
  \item Explicit mixed-size spreading targets
  derived from the die geometry and object sizes,
  which remain valid under extreme size variation (Section~\ref{subsec:mixedsize_moments_method_v4});

  \item Stable constraint construction that accounts for fixed objects, centering, and utilization-aware rescaling, preventing pathological spreading in early iterations and preserving feasibility (Sections~\ref{subsec:center_C_method_v4}--\ref{subsec:C_scale_v4});

  \item A Stiefel transformation that makes the QCQP constraint geometry explicit through metric equalization and projection, enabling scalable Riemannian optimization by SSM with second-order steps. (Section~\ref{subsec:stiefel_transform_solver});

  \item Geometry-consistent sparse modeling for multi-pin nets by
  integrating an update loop with an angle-sorted cycle sparsification to keep the Laplacian sparse while maintaining net geometry during initialization (Section~\ref{subsec:sparsification});

  \item Evaluation of our initializer as a drop-in replacement for common heuristic initializations, with downstream weighted HPWL (wHPWL) improvements across RePlAce~\cite{ChengKKW19_RePlAce}, DREAMPlace~\cite{LinDLRKP19_DREAMPlace}, and AutoDMP~\cite{AutoDMP23_ISPD} on 20 benchmarks (Section~\ref{sec:experiments}).
\end{itemize}

\paragraph*{Relation to prior work} Chen et al.~\cite{ChenCH24_StiefelSSM_arXiv}
develop the abstract Stiefel-SSM framework and its convergence
theory for quadratic objectives on $St(n,r)$. This paper uses
that solver framework after deriving a placement-specific Stiefel
problem from a mixed-size QCQP. The required modeling steps
include mixed-size spreading-target construction, fixed-object
and utilization corrections, metric equalization of the movable-
variable QCQP, and geometry-updated sparse Laplacian modeling.
Compared with the sphere-constrained initializer of
Chen et al.~\cite{ChenCCHLW23_ISPD}, the present formulation
removes the averaged-size spreading simplification, incorporates
fixed objects and centering directly into the movable-variable
constraints, and uses the Stiefel-SSM solver with periodic
angle-sorted-cycle updates.


The remainder of the paper is organized as follows.
Section~\ref{sec:motivation} discusses why mixed-size initialization is challenging and identifies gaps in existing approaches.
Section~\ref{sec:background} reviews placement notation, wirelength metrics, and Stiefel manifold basics.
Section~\ref{sec:method} presents our mixed-size QCQP formulation, Stiefel transformation, and the Stiefel sequential subspace solver with update-loop sparsification.
Section~\ref{sec:experiments} reports experimental results on both industry and academic benchmarks, and Section~\ref{sec:conclusion} concludes.


\section{Motivation}
\label{sec:motivation}

\subsection{Challenges of Mixed-Size Initialization}
\label{subsec:mixedsize_challenges}

Mixed-size initialization is harder than its
standard-cell-only counterpart for three
related reasons.

\subsubsection{Area-Aware Modeling}
Mixed-size instances contain macros whose
footprints differ by orders of magnitude from
standard cells.
Moreover, macros themselves vary widely in
size, which further complicates the
construction of spreading constraints.
Ignoring object sizes causes severe overlap or
collapse. Overcompensating causes over-expansion
beyond the die.
Effective initialization therefore requires
\emph{size-aware spreading targets} that account
for each object's individual
dimensions~\cite{LuZhuangChenEtAl15_TCAD_ePlaceMS,
YanViswanathanChu09_DAC_MixedSizeComplexities}.

\subsubsection{Initialization Sensitivity}
Global placement objectives are highly nonconvex,
and under finite iteration budgets different
initializations converge to different basins,
producing different wHPWL outcomes (see Section~\ref{subsec:hpwl_quad} for the formal definition).
This motivates \emph{quality-focused}
initializers that provide a well-spread,
structure-aware initial placement.

\subsubsection{Normalization Difficulty}
Large area disparities create ill-conditioned
constraints, and fixed objects shift the free
region and bias the centroid.
A robust mixed-size initializer requires careful
normalization, centering, and feasibility checks
to avoid pathologies such as over-expansion or
constraint
indefiniteness (Section~\ref{subsec:C_scale_v4}
presents our solution).

\subsection{Gaps in Existing QCQP Initialization}
\label{subsec:qcqp_gaps}

A prominent approach to constrained spectral initialization is to minimize a quadratic Laplacian wirelength approximation subject to first-order centroid and second-order spreading equalities, i.e., a QCQP.
Despite strong empirical results, the mixed-size QCQP initialization still faces two practical difficulties.

\subsubsection{Spreading-Target Construction Fragility}
The most fragile part of QCQP initialization is often not the Laplacian objective but the \emph{spreading-target construction}: building a feasible, well-conditioned constraint set when fixed objects, mixed sizes, and nontrivial utilization are present.
Even when prior work demonstrates mixed-size experiments, the derivation and implementation of spreading targets are frequently simplified, leaving the constraint constants and matrix construction inadequate for varying macro sizes~\cite{ChenCCHLW23_ISPD}.

\subsubsection{Quadratic Equality Constraints}
Large-scale QCQP solvers for placement initialization can be numerically delicate, because quadratic equality constraints define a low-dimensional feasible set and amplify scaling issues in the presence of large area disparities~\cite{YanViswanathanChu09_DAC_MixedSizeComplexities}.
These spreading-target construction issues are especially pronounced in industry-scale designs where a small number of macros dominate area and connectivity while interacting with a massive standard-cell population and multiple classes of fixed objects.

\subsubsection{Stiefel Formulation}
The two difficulties above share a common geometric root: the quadratic spreading equalities that enforce
global spread define a \emph{generalized orthogonality constraint} $\bm X^\top \bm G \bm X = \bm C$ in the space of placement coordinates. After a metric-equalization step that absorbs the area weights $\bm G$ and the spreading targets $\bm C$, this constraint reduces to the standard orthonormality condition $\bm{\tilde X}^{\top}\bm{\tilde X}=\bm I$, i.e., the feasible set becomes a Stiefel manifold.
Making this structure explicit brings three practical benefits.
First, feasibility is maintained \emph{by construction} at every iteration, eliminating the fragile penalty-based or projection-based enforcement needed by generic QCQP
solvers.
Second, the smooth manifold geometry enables Riemannian
Newton conjugate-gradient (Newton-CG) updates on tangent spaces.
By incorporating local curvature information, these updates can
improve convergence behavior under the ill-conditioning caused by
large area disparities.
Third, the Stiefel viewpoint decouples the centroid constraint from the spreading constraint via a rank-one projector, simplifying the overall problem without sacrificing sparsity.
Section~\ref{sec:method} presents the details.


\section{Background and Preliminaries}
\label{sec:background}


\begin{table}[htbp]
\centering
\caption{Summary of key notations}
\label{tab:notation}
\vspace{-4pt}
\begin{footnotesize}
\renewcommand{\arraystretch}{1.1}
\begin{tabular}{@{}ll@{}}
\toprule
\textbf{Symbol} & \textbf{Description} \\
\midrule
$\mathcal{H}=(\mathcal{V},\mathcal{N})$ & Netlist hypergraph \\
$\mathcal{V},\;\mathcal{N}$ & Set of objects, set of nets \\
$\mathcal{F},\;\mathcal{M}$ & Fixed and movable object sets \\
$n,\;n_{\mathrm{free}}$ & Total and movable object counts \\
$L_d,\,H_d;\;\ell_i,\,h_i$ & Die / object width and height \\
$w_i;\;\bm G$ & Area weight; diagonal weight matrix \\
$\bm X\in\mathbb{R}^{n\times 2}$ & Coordinate matrix \\
$\bm L,\;\bm A,\;\bm D$ & Graph Laplacian, adjacency and degree \\
$\bm C \in \mathbb{R}^{2\times 2}$ & Spreading constraint matrix \\
$c_1,\,c_2,\,c_3 \in \mathbb{R}$ & Mixed-size spreading-target constants \\
$\rho$ & Utilization ratio for $\bm C$ rescaling \\
$\widetilde{\bm L}$ & Equalized Laplacian, Eq.~\eqref{eq:equalization_v4} \\
$\bm P$ & Rank-one centroid projector, Eq.~\eqref{eq:P_v4} \\
$\widehat{\bm L};\;\bm B$ & Projected Laplacian and linear term, Eq.~\eqref{eq:AB_v4} \\
$\widehat{\bm L}_e$ & Spectrally shifted $\widehat{\bm L}$, Eq.~\eqref{eq:spectral_shift_v4} \\
$\bm V_k;\;\bm{\check X}$ & SSM subspace basis; reduced variable \\
$\bar{\bm L}_k,\;\bar{\bm B}_k$ & Reduced operator and linear term on $\bm V_k$ \\
$\bm\Lambda$ & Stiefel Lagrange multiplier (symmetric) \\
$\mathrm{St}(n,r)$ & Stiefel manifold $\{\bm X\in\mathbb{R}^{n\times r}:\bm X^{\!\top}\bm X=\bm I_r\}$ \\
$\bm{\tilde X}\in\mathrm{St}(n_{\mathrm{free}},2)$ & Stiefel variable (post-equalization) \\
\bottomrule
\end{tabular}
\end{footnotesize}
\end{table}

This section reviews the notation and background needed for the proposed method.
We introduce mixed-size placement conventions, the wirelength metrics used for evaluation, and basic definitions for the Stiefel manifold. Table~\ref{tab:notation} summarizes the notation used throughout Sections~\ref{sec:background} and ~\ref{sec:method}.

\subsection{Mixed-Size Placement Notation}
\label{subsec:prelim_mixedsize}

A placed circuit is represented by a netlist hypergraph $\mathcal{H}=(\mathcal{V},\mathcal{N})$, where $\mathcal{V}$ is the set of placement objects and $\mathcal{N}$ is the set of nets connecting multiple pins.
Mixed-size placement refers to instances in which $\mathcal{V}$ contains both \emph{standard cells} and much larger \emph{macros}, often with footprint disparities of orders of magnitude~\cite{MarkovHuKim15_ProcIEEE_PlacementSurvey,LuZhuangChenEtAl15_TCAD_ePlaceMS,YanViswanathanChu09_DAC_MixedSizeComplexities}.

\subsubsection{Die and Coordinate Variables}
We denote the placement region (die) by an axis-aligned rectangle of width $L_d$ and height $H_d$.
For $n=|\mathcal{V}|$ objects, let
 $\bm{x}, \bm{y}\in \mathbb{R}^{n}$ denote the vectors of object center coordinates.
We write
\[
\bm{X}=[\bm{x}\ \ \bm{y}]\in\mathbb{R}^{n\times 2},
\]
so that row $i$ of $\bm X$ is $[x_i\ \ y_i]$, the center coordinates of object $i$.
Initialization is a \emph{global} placement stage: the coordinates are continuous and overlaps are allowed, with the goal of producing a well-spread initial placement for downstream density-driven refinement and legalization.

\subsubsection{Fixed and Movable Objects}
We partition objects into a \emph{fixed} set $\mathcal{F}$ (preplaced blocks and I/O terminals) and a \emph{movable} set $\mathcal{M}$ (standard cells and macros), with
\[
n_{\mathrm{fixed}} = |\mathcal{F}|,\qquad
n_{\mathrm{free}}  = |\mathcal{M}|,\qquad
n = n_{\mathrm{fixed}} + n_{\mathrm{free}}.
\]
Accordingly, we split the coordinate vectors and other related vectors into fixed and movable blocks,
\[
\bm x=\begin{bmatrix}\bm x_F\\\bm x_M\end{bmatrix},\quad
\bm y=\begin{bmatrix}\bm y_F\\\bm y_M\end{bmatrix},\quad
\bm X=\begin{bmatrix}\bm X_F\\\bm X_M\end{bmatrix},
\]

where $(\bm x_F,\bm y_F)$ are the vectors for the fixed objects and $(\bm x_M,\bm y_M)$ are the vectors for the movable objects, which are initialized by the proposed method.

\subsubsection{Object Sizes and Area Weights}

Each object $i$ has width $\ell_i$ and height $h_i$. We denote its area weight by
$w_i=\ell_i h_i,$
and assemble the area-weight vector as
$\bm{w}\in\mathbb{R}^n.$
The corresponding diagonal area matrix is
\[
\bm{G}=\operatorname{diag}(\bm{w}).
\]
With the fixed/movable partition,
\[
\bm{w}=
\begin{bmatrix}
\bm{w}_F\\
\bm{w}_M
\end{bmatrix},
\qquad
\bm{G}=
\begin{bmatrix}
\bm{G}_F & \bm{0}\\
\bm{0} & \bm{G}_M
\end{bmatrix},
\]
where
\[
\bm{G}_F=\operatorname{diag}(\bm{w}_F),
\qquad
\bm{G}_M=\operatorname{diag}(\bm{w}_M).
\]

\subsection{HPWL and Quadratic Wirelength
  Approximation}
\label{subsec:hpwl_quad}

For a net $e\in\mathcal{N}$ with pins
$\mathcal{V}(e)\subseteq\mathcal{V}$, HPWL is
\begin{equation}
\label{eq:hpwl_def}
\mathrm{HPWL}(e)=
\Big(\max_{i\in\mathcal{V}(e)} x_i
  -\min_{i\in\mathcal{V}(e)} x_i\Big)
+
\Big(\max_{i\in\mathcal{V}(e)} y_i
  -\min_{i\in\mathcal{V}(e)} y_i\Big),
\end{equation}
and wHPWL aggregates nets with
optional weights $\{\omega_e\}$ to reflect their degree:
\begin{equation}
\label{eq:whpwl_def}
\mathrm{wHPWL}(X)
  =\sum_{e\in\mathcal{N}}
    \omega_e\,\mathrm{HPWL}(e).
\end{equation}
We report wHPWL as the main quality metric.

For initialization, we use a quadratic Laplacian energy. We first reduce
the netlist hypergraph to a sparse weighted graph with adjacency $\bm A$.
The nonzero off-diagonal entries of $\bm A$ correspond to the pairwise
edges selected from each net; for multi-pin nets, these edges are formed
by the angle-sorted cycle model in Section~\ref{subsec:sparsification}.
The graph Laplacian is
\[
\bm L=\bm D-\bm A,
\]
where $\bm D$ is the corresponding weighted degree matrix. The quadratic
objective is
\begin{equation}
\label{eq:laplacian_energy}
\Phi(\bm x,\bm y)=\bm{x}^\top \bm L\bm{x}
  +\bm{y}^\top \bm L\bm{y}
  =\mathrm{tr}\!\big(\bm{X}^\top \bm L\bm{X}\big).
\end{equation}
This sparse quadratic form supports fast iterative solvers and underpins
the QCQP initialization in Section~\ref{subsec:qcqp_formulation}. It
captures connectivity through squared pairwise distances on the reduced
graph.
\subsection{Stiefel Manifold}
\label{subsec:stiefel_prelim}

For integers $r\le n$, the \emph{Stiefel manifold} is
the set (endowed with a manifold structure) of all
matrices whose columns are orthonormal, that is,
\begin{equation}
\label{eq:stiefel_def}
\mathrm{St}(n,r)
  =\{\bm X\in\mathbb{R}^{n\times r}:
     \bm X^\top \bm X=\bm I_r\}.
\end{equation}

We next recall the Riemannian tools~\cite{AbsilMahonySepulchre08,EdelmanAriasSmith98} on the
Stiefel manifold (Eq.~\eqref{eq:stiefel_def}) needed
for our proposed optimization method.
The central idea is to optimize directly on the manifold that represents the feasible set. Instead of enforcing the quadratic spreading equalities through penalty terms or generic Euclidean projections, we change variables so that the feasible set \emph{becomes} the search space itself.
In our setting, the quadratic spreading equalities of the QCQP, after metric equalization, reduce to the orthonormality constraint $\bm{\tilde X}^\top \bm{\tilde X} = \bm I$, and the search space becomes the Stiefel manifold (Eq.~\eqref{eq:stiefel_def}).
This enables the use of Riemannian optimization tools that maintain feasibility by construction.

\subsubsection{Tangent Space and Projection}
The tangent space at
$\bm X\in\mathrm{St}(n,r)$ is
\begin{equation}
\label{eq:stiefel_tangent}
T_{\bm X}\mathrm{St}(n,r)
  =\{\bm Z\in\mathbb{R}^{n\times r}:
     \bm X^\top \bm Z+\bm Z^\top \bm X=\bm 0\}.
\end{equation}
For any $\bm U\in\mathbb{R}^{n\times r}$, its
orthogonal projection onto
$T_{\bm X}\mathrm{St}(n,r)$ is given by
\begin{equation}
\begin{gathered}
\label{eq:stiefel_proj}
\mathrm{Proj}_{\bm X}(\bm U)
  =\bm U-\bm X\,\mathrm{sym}(\bm X^\top \bm U),
\\
\mathrm{sym}(\bm M)=\tfrac12(\bm M+\bm M^\top).
\end{gathered}
\end{equation}
The symmetric-part operator $\mathrm{sym}(\cdot)$, equivalently $(\cdot)_{\mathrm{sym}}$, returns $\tfrac{1}{2}(\bm M+\bm M^\top)$ for a square matrix $\bm M$.

\subsubsection{Riemannian Gradient}
For a smooth $f$ restricted to
$\mathrm{St}(n,r)$,
\begin{equation}
\label{eq:riem_grad_prelim}
\mathrm{grad}\,f(\bm X)
  = \mathrm{Proj}_{\bm X}(\nabla f(\bm X)).
\end{equation}
A first-order critical point on the Stiefel manifold satisfies
$\nabla f(\bm X)=\bm X\bm \Lambda$ for some symmetric
$\bm \Lambda$~\cite{EdelmanAriasSmith98,
AbsilMahonySepulchre08}. 

\subsubsection{Retraction}
The polar (metric-projection) retraction maps tangent vectors back to
the manifold:
\begin{equation}
\label{eq:polar_retraction}
\mathrm{Retr}_{\bm X}(\bm Z)=(\bm X+\bm Z)\big((\bm X+\bm Z)^\top(\bm X+\bm Z)\big)^{-1/2}=\bm U \bm V^{\!\top},
\end{equation}
where $\bm U \bm \Sigma \bm V^{\!\top}$ is the thin singular-value decomposition of $\bm X+\bm Z$~\cite{AbsilMahonySepulchre08,ChenCH24_StiefelSSM_arXiv}.

\subsubsection{Rank-One Projector}
For a nonzero $\bm p\in\mathbb{R}^n$,
\begin{equation}
\label{eq:rank1_projector_prelim}
\bm P = \bm I - \frac{\bm p \bm p^\top}{\bm p^\top \bm p}
\end{equation}
enforces $\bm p^\top \bm X=\bm 0^\top$ for any $\bm X$ in its range
while preserving sparsity up to a rank-one
correction.

Section~\ref{subsec:stiefel_transform_solver} describes our solver, which combines these Riemannian tools with SSM~\cite{ChenCH24_StiefelSSM_arXiv} tailored to the Stiefel-constrained QCQP, using the manifold's smooth geometry for scalable Riemannian Newton-CG optimization.


\section{Proposed Method}
\label{sec:method}

We address the challenges and gaps of
Section~\ref{sec:motivation} by making the
geometric structure of the QCQP constraints
explicit.
After metric equalization, the quadratic spreading
equalities become an orthonormality condition on
a Stiefel manifold, enabling the use of
Riemannian optimization with
feasibility-preserving steps and stable
second-order solvers.
Combined with careful mixed-size spreading-target
construction and utilization-aware rescaling,
this provides a robust initialization framework
for mixed-size placement.

We propose a mixed-size placement initializer
that is formulated as a
constrained quadratic program with first- and second-order constraints. We make its
quadratic equality constraints
\emph{explicitly geometric} via a Stiefel
manifold transformation, and solve the resulting
problem with a Riemannian sequential subspace
method.
The output is a continuous initial placement of
center coordinates for movable objects that can be
passed as a drop-in initialization to downstream placers along with the coordinates of the fixed objects.


\subsection{Mixed-Size QCQP Formulation}
\label{subsec:problem_construction}

\subsubsection{QCQP Formulation}
\label{subsec:qcqp_formulation}

We start from the constrained QCQP
formulation of~\cite{ChenCCHLW23_ISPD} and
reformulate the constraints to expose their
geometric structure, minimizing the quadratic
Laplacian energy (Eq.~\eqref{eq:laplacian_energy})
subject to global first- and second-order
constraints under area weights.

\paragraph{First- and second-order constraints}
A representative constrained spectral initializer is
\begin{equation}
\label{eq:qcqp_template_prelim}
\begin{aligned}
\min_{\bm x,\bm y\in\mathbb{R}^n}\quad & \bm x^{\top}\bm L\bm x + \bm y^{\top}\bm L\bm y\\
\text{s.t.}\quad
& \bm w^{\top}\bm x = c_x,\qquad \bm w^{\top}\bm y = c_y,\\
& \bm x^{\top}\bm G\bm x = c_1,\;\;\;\;\bm y^{\top}\bm G\bm y = c_2,\;\;\;\; \bm x^{\top}\bm G\bm y = c_3 .
\end{aligned}
\end{equation}
The first-order constraints are linear and fix the area-weighted centroid, removing the global translation ambiguity inherent in Laplacian energies.
The second-order constraints are quadratic and control the global spread and orientation of the embedding.

\paragraph{Compact matrix form}
By stacking $\bm X=[\bm x\ \bm y]$ and $\bm c_0=[c_x\ c_y]^\top$, and defining
\[
\bm C_{\mathrm{raw}}=\begin{bmatrix}c_1&c_3\\c_3&c_2\end{bmatrix},
\]
the constraints can be written compactly as
\begin{equation}
\label{eq:moment_matrix_constraint}
\bm w^\top \bm X=\bm c_0^\top,\qquad \bm X^\top \bm G \bm X= \bm C_{\mathrm{raw}}.
\end{equation}
The constraint matrix $\bm C$ specifies target spreading statistics of the placement distribution: choosing $\bm C$ to match the geometry of the available placement region enforces physically meaningful spreading.

\paragraph{Reduction to movable variables}
While fixed objects contribute to the objective through net connections, they are not decision variables.
Partitioning $\bm L$ conformally with $(\bm X_F,\bm X_M)$ and dropping constants independent of $\bm X_M$ yields the reduced objective
\begin{equation}
\begin{gathered}
\label{eq:trace_fixed_reduction}
\min_{\bm X_M\in\mathbb{R}^{n_{\mathrm{free}}\times 2}}\ 
\mathrm{tr}(\bm X_M^\top \bm L_{MM}\bm X_M)+2\,\mathrm{tr}(\bm E^\top \bm X_M),
\\
\bm E=\bm L_{MF}\bm X_F.
\end{gathered}
\end{equation}
The spreading constraints are similarly restricted to the movable blocks by replacing $(\bm w,\bm G,\bm X)$ with $(\bm w_M,\bm G_M,\bm X_M)$ and adjusting the right-hand sides to account for the fixed contribution~\cite{ChenCCHLW23_ISPD}.\\

\subsubsection{Mixed-Size Spreading Targets}
\label{subsec:mixedsize_moments_method_v4}

The QCQP
formulation (Eq.~\eqref{eq:qcqp_template_prelim})
requires target values
$(c_x,c_y,c_1,c_2,c_3)$.
We set the coordinate origin at the die center,
hence $c_x=c_y=0$.
We also set $c_3=0$, which enforces axis-aligned spreading with zero cross-correlation between the $x$ and $y$ coordinates.
This is appropriate for centered rectangular dies with axis-aligned objects. Extending the formulation to non-axis-aligned placement regions or rotated objects would require computing the full second-moment matrix, including the $ x y$ term when it is nonzero. This extension is left as future work.
We then compute mixed-size
targets $c_1$ and $c_2$ that account for varying rectangle sizes as follows.

Let the die be the axis-aligned rectangle $\Omega_{\mathrm{die}}=[-L_d/2,L_d/2]\times[-H_d/2,H_d/2]$.
For a uniform distribution over $\Omega_{\mathrm{die}}$, the second moments about the origin follow by direct integration. Along the $x$-axis,
\begin{align*}
\int_{\Omega_{\mathrm{die}}} x^2\,\mathrm{d}A
&= H_d \int_{-L_d/2}^{L_d/2}\! x^2\,\mathrm{d}x \\
&= H_d\!\left[\frac{x^3}{3}\right]_{-L_d/2}^{L_d/2} \\
&= \frac{2H_d}{3}\!\left(\frac{L_d}{2}\right)^{\!3} \\
&= \frac{H_d L_d^3}{12}.
\end{align*}
Symmetric reasoning along the $y$-axis, together with $\int xy\,\mathrm{d}A=0$ by symmetry of the centered rectangle, gives
\begin{equation}
\begin{gathered}
\label{eq:die_moments}
\int_{\Omega_{\mathrm{die}}} x^2\,\mathrm{d}A = \frac{H_d L_d^3}{12},
\;
\int_{\Omega_{\mathrm{die}}} y^2\,\mathrm{d}A = \frac{L_d H_d^3}{12},
\\
\int_{\Omega_{\mathrm{die}}} xy\,\mathrm{d}A = 0.
\end{gathered}
\end{equation}
We relate these continuous target values to discrete placement
variables as follows. For an axis-aligned rectangle object $i$
with center $(x_i,y_i)$, width $\ell_i$, height $h_i$, area
$w_i=\ell_i h_i$, and occupied region $\Omega_i$, the
$x$-second-moment contribution decomposes as
\[
\int_{\Omega_i} x^2\,dA
=
\underbrace{w_i x_i^2}_{\text{center term}}
+
\underbrace{I_{x,i}}_{\text{internal term}},
\qquad
I_{x,i}=\frac{h_i\ell_i^3}{12}.
\]
Similarly,
\[
\int_{\Omega_i} y^2\,dA
=
w_i y_i^2 + I_{y,i},
\qquad
I_{y,i}=\frac{\ell_i h_i^3}{12}.
\]
Because each rectangle is axis-aligned and referenced to its
own center, the internal cross term vanishes: $I_{xy,i}=0$.

Therefore, if the union of all rectangles is intended to match the die-level target values, the \emph{center terms} should satisfy
\[
\sum_{i=1}^{n} w_i x_i^2
=
\frac{H_d L_d^3}{12} - \sum_{i=1}^{n} I_{x,i},
\]
\[
\sum_{i=1}^{n} w_i y_i^2
=
\frac{L_d H_d^3}{12} - \sum_{i=1}^{n} I_{y,i}.
\]
Substituting Eq.~\eqref{eq:die_moments} and the internal-contribution formulas gives the mixed-size spreading targets
\begin{equation}
\label{eq:moment_constants_v4}
\begin{aligned}
c_1&=\frac{1}{12}\!\left(H_d L_d^{3}-\sum_{i=1}^{n} h_i \,\ell_i^{3}\right),\\
c_2&=\frac{1}{12}\!\left(L_d H_d^{3}-\sum_{i=1}^{n} \ell_i \,h_i^{3}\right),\\
c_3&=0,
\qquad c_x=c_y=0.
\end{aligned}
\end{equation}
The expressions in Eq.~\eqref{eq:moment_constants_v4} give closed-form targets, require no numerical integration, and yield size-aware spreading targets because each object's internal contribution is subtracted with its own width $\ell_i$ and height $h_i$.
The sphere-constrained formulation of ~\cite{ChenCCHLW23_ISPD} uses a simplified spreading-target construction based on averaged object dimensions. Here, each object's width and height are subtracted individually.
Our formulation computes size-aware spreading targets that subtract each object's internal contribution using its own width and height, remaining stable even when macro sizes vary widely.\\

\subsubsection{Centering and Constraint Matrix
  Construction}
\label{subsec:center_C_method_v4}

The most delicate part of QCQP initialization in
mixed-size settings is the stable construction of
a feasible, well-conditioned spreading
constraint under fixed objects and nontrivial
utilization.
We present the centering and the resulting
right-hand side in a form that matches the
implementation.
\\
\paragraph{Centering transformation}

The centroid constraints in Eq.~\eqref{eq:qcqp_template_prelim}, restricted to the movable block, read as
\[
\bm w_M^\top \bm x_M = c_x-\bm w_F^\top \bm x_F,
\qquad
\bm w_M^\top \bm y_M = c_y-\bm w_F^\top \bm y_F.
\]
Let $s=\mathbf{1}^\top \bm w_M$ and define the weighted shifts
\begin{equation}
\label{eq:deltas_v4}
\Delta_x=\frac{c_x-\bm w_F^\top \bm x_F}{s},
\qquad
\Delta_y=\frac{c_y-\bm w_F^\top \bm y_F}{s}.
\end{equation}
We recenter the movable coordinates as
\begin{equation}
\begin{gathered}
\label{eq:center_vars_v4}
\bm{\tilde x}_M = \bm x_M-\Delta_x\mathbf{1},
\qquad
\bm{\tilde y}_M = \bm y_M-\Delta_y\mathbf{1},
\\
\bm{\tilde X}_M=[\bm{\tilde x}_M\ \bm{\tilde y}_M],
\end{gathered}
\end{equation}
so that $\bm w_M^\top  \bm{\tilde x}_M=0$ and $\bm w_M^\top \bm{\tilde y}_M=0$ hold identically.

Substituting $\bm X_M=\bm{\tilde X}_M+\mathbf{1}[\Delta_x\ \Delta_y]$ into Eq.~\eqref{eq:trace_fixed_reduction} preserves the quadratic term and produces a shifted linear term.
The centered objective is
\begin{equation}
\label{eq:centered_trace_v4}
\min_{\tilde X_M\in\mathbb{R}^{n_{\mathrm{free}}\times 2}}
\quad
\mathrm{tr}(\bm{\tilde X}_M^\top \bm L_{MM}\bm{\tilde X}_M) + 2\,\mathrm{tr}(\bm E_0^\top \bm{\tilde X}_M),
\end{equation}
with the explicit expression
\begin{equation}
\label{eq:E0_v4}
\bm E_0=
\Big[\,\bm b + \Delta_x\,\bm L_{MM}\mathbf{1}\;\;\;\; \bm d + \Delta_y\,\bm L_{MM}\mathbf{1}\,\Big],
\end{equation}
where $\bm E=[\bm b\ \bm d]$ with $\bm b=\bm L_{MF}\bm x_F$ and $\bm d=\bm L_{MF}\bm y_F$.

\paragraph{Constraint matrix}

Let $\bm C_{\mathrm{raw}}$ be the $2\times 2$ target spreading matrix formed from Eq.~\eqref{eq:moment_constants_v4}:
\[
\bm C_{\mathrm{raw}}=
\begin{bmatrix}
c_1 & c_3\\
c_3 & c_2
\end{bmatrix}.
\]

The spreading constraint on the full variables is $\bm X^\top \mathrm{diag}(\bm w)\,\bm X=\bm C_{\mathrm{raw}}$.
Restricting this constraint to the movable block and substituting the centered variables (Eq.~\eqref{eq:center_vars_v4}) yields
\begin{equation}
\label{eq:centered_moment_constraint_v4}
\bm w_M^\top \bm{\tilde X}_M=\bm{0}^\top,
\qquad
\bm{\tilde X}_M^\top \bm G_M \bm{\tilde X}_M = \bm C,
\end{equation}
where the right-hand side $\bm C$ is
\begin{equation}
\begin{gathered}
\label{eq:C_construct_v4}
\bm C
=
\bm C_{\mathrm{raw}}
\;-\;
\bm X_F^\top \bm G_F \bm X_F
\;-\;
\frac{1}{s}\,\tilde{\bm c}_0 \tilde{\bm c}_0^{\!\top},
\\
\tilde{\bm c}_0=
\begin{bmatrix}
c_x-\bm w_F^{\!\top} \bm x_F\\
c_y-\bm w_F^{\!\top} \bm y_F
\end{bmatrix}.
\end{gathered}
\end{equation}

The first subtraction removes the  \emph{center contributions} of the fixed objects from the global target.
The rank-one term $\frac{1}{s}\,\tilde{\bm c}_0 \tilde{\bm c}_0^{\!\top}$ is the centering correction induced by shifting the movable coordinates so that $\bm w_M^{\!\top} \bm{\tilde X}_M=\bm{0}^\top$.
Here $\tilde{\bm c}_0$ is the fixed-object-adjusted centroid target. It is distinct from the raw target $\bm c_0=[c_x,c_y]^{\!\top}$ defined earlier.\\

\subsubsection{Utilization-Aware Rescaling}
\label{subsec:C_scale_v4}

In mixed-size benchmarks, the free region available to movable objects can differ significantly from the total movable area due to fixed obstacles and whitespace.
If one enforces $\bm C$ from Eq.~\eqref{eq:C_construct_v4} directly, the initializer can over-spread the movable coordinates when utilization is below one, leading to extreme excursions that subsequent iterations cannot easily correct.
We mitigate this by rescaling $\bm C$ according to the ratio between movable area and free-region area.

We use the following area quantities:
\[
A_{\mathrm{mov}}=\sum_{i\in\mathcal{M}} w_i,\;
A_{\mathrm{fix}}=\sum_{i\in\mathcal{F}} w_i,\;
\]
\[
A_{\mathrm{die}}=L_d H_d,\;
A_{\mathrm{free}}=A_{\mathrm{die}}-A_{\mathrm{fix}}.
\]
We assume $A_{\mathrm{mov}}\le A_{\mathrm{free}}$ so that the movable objects fit within the free region. Over-utilized instances are outside the present scope.
Define
\begin{equation}
\label{eq:C_scale_v4_eq}
\rho=\frac{A_{\mathrm{mov}}}{A_{\mathrm{free}}}\in(0,1],
\qquad
\bm C \leftarrow \rho^{2}\,\bm C.
\end{equation}
If movable objects uniformly cover a sub-region of $\Omega_{\mathrm{free}}$ similar in shape to $\Omega_{\mathrm{free}}$ with area $A_{\mathrm{mov}}$, the linear factor is $\sqrt{\rho}$.
Under this similarity assumption, all linear dimensions scale by
$\sqrt{\rho}$. Because a second moment integrates a degree-two
coordinate term over area, each entry of $\bm C$ scales by
$(\sqrt{\rho})^2 \rho = \rho^2$, giving the rescaling
$\bm C \leftarrow \rho^2 \bm C$.
Next, we numerically verify that $\bm C$ is symmetric positive definite (SPD). If the check fails, the algorithm aborts. This check did not fail for any benchmark in our evaluation.


\subsection{Stiefel Transformation and Solver}
\label{subsec:stiefel_transform_solver}

We apply Riemannian tools on the
Stiefel manifold (Eq.~\eqref{eq:stiefel_def}) from Section~\ref{subsec:stiefel_prelim} to present the
transformation and our solver.

\subsubsection{Metric Equalization and Stiefel
  Transformation}
\label{subsec:stiefel_transform_v4}

We now convert the quadratic equalities
into a Stiefel constraint by
converting Eq.~\eqref{eq:centered_trace_v4} with
constraint Eq.~\eqref{eq:centered_moment_constraint_v4}
into an optimization on the Stiefel manifold.
Geometrically, the constraint $\bm{\tilde X}_M^\top \bm G_M \bm{\tilde X}_M = \bm C$ is a generalized orthogonality condition whose geometry is determined by the area weights $\bm G_M$ and the spreading targets $\bm C$.
The metric equalization below absorbs these weights into a change of variables, converting the constraint to the standard orthonormality condition on the Stiefel manifold $\mathrm{St}(n_{\mathrm{free}},2)$.
This step bridges QCQP-based initialization and Riemannian optimization.
The transformation is well-defined whenever $\bm G_M$ and $\bm C$ are symmetric positive definite, which holds because all movable-object areas satisfy $w_i > 0$ and because $\bm C$ is verified to be SPD after rescaling in Section~\ref{subsec:C_scale_v4}.

\paragraph{Equalization of the quadratic constraint}
Define the equalized operator and linear term
\begin{equation}
\label{eq:equalization_v4}
\tilde{\bm L} = \bm G_M^{-1/2}\bm L_{MM}\bm G_M^{-1/2},
\qquad
\widetilde{\bm{E}}_0 = \big(\bm{C}^{1/2}\bm E_0^{\top} \bm{G}_M^{-1/2}\big)^\top,
\end{equation}
and introduce the Stiefel variable
\begin{equation}
\label{eq:stiefel_var_v4}
\bm{\tilde X} = \bm G_M^{1/2}\bm{\tilde X}_M\,\bm C^{-1/2}\in\mathbb{R}^{n_{\mathrm{free}}\times 2}.
\end{equation}
Note that $\bm{\tilde X}$ is distinct from the physical coordinate matrix $\bm X$ of Section~\ref{subsec:prelim_mixedsize}. The relation is given by Eq.~\eqref{eq:stiefel_var_v4}.
Then the quadratic equality $\bm{\tilde X}_M^\top \bm G_M\bm{\tilde X}_M=\bm C$ becomes $\bm{\tilde X}^\top \bm{\tilde X}=\bm I_2$, i.e., $\bm{\tilde X}\in\mathrm{St}(n_{\mathrm{free}},2)$ (Eq.~\eqref{eq:stiefel_def}).

\paragraph{Centroid projector}
The centered centroid constraint $\bm w_M^\top \bm{\tilde X}_M=\bm{0}^\top$ becomes $\bm{\tilde w}^\top \bm{\tilde X}=\bm{0}^\top$ with $\bm{\tilde w}=\bm G_M^{-1/2}\bm w_M$.
We enforce this constraint implicitly by the rank-one projector (Eq.~\eqref{eq:rank1_projector_prelim}):
\begin{equation}
\label{eq:P_v4}
\bm P = \bm I - \frac{\bm{\tilde w}\bm{\tilde w}^\top}{\bm{\tilde w}^\top \bm{\tilde w}}.
\end{equation}
We then form the projected equalized Laplacian and projected linear term
\begin{equation}
\label{eq:AB_v4}
\widehat{\bm L} = \bm P\,\widetilde{\bm L}\,\bm P,
\qquad
\bm B = \bm P\,\widetilde{\bm{E}}_0,
\end{equation}
and obtain the transformed Stiefel problem
\begin{equation}
\label{eq:stiefel_obj_v4}
\min_{\bm{\tilde X}\in\mathrm{St}(n_{\mathrm{free}},2)}
\quad
f(\bm{\tilde X})=\mathrm{tr}\!\big(\bm{\tilde X}^\top \widehat{\bm L}\, \bm{\tilde X}\,\bm C\big)+2\,\mathrm{tr}(\bm B^\top \bm{\tilde X}).
\end{equation}
Because $\bm P$ differs from the identity by rank one, Eq.~\eqref{eq:stiefel_obj_v4} preserves the sparsity of $\bm L_{MM}$ up to low-rank corrections.
All required operator applications can therefore be implemented using sparse matrix--vector products and rank-one updates.
By construction, the initial iterate and every subspace-basis column lie in $\mathrm{range}(\bm P)$, so the trajectory remains in $\mathrm{St}(n_{\mathrm{free}},2)\cap\mathrm{range}(\bm P)$ throughout the optimization.

\paragraph{Back-mapping to physical coordinates}
Given a solution $\bm{\tilde X}$ of Eq.~\eqref{eq:stiefel_obj_v4}, we recover the centered movable coordinates by inverting Eq.~\eqref{eq:stiefel_var_v4}:
\[
\bm{\tilde X}_M = \bm G_M^{-1/2}\bm{\tilde X}\,\bm C^{1/2}.
\]
We then undo centering via Eq.~\eqref{eq:center_vars_v4}:
\[
\bm X_M = \bm{\tilde X}_M + \mathbf{1}[\Delta_x\ \Delta_y],
\]
and concatenate $(\bm X_F,\bm X_M)$ to obtain a full initial placement.\\

\subsubsection{Stiefel-SSM Solver}
\label{subsec:ssm_update_v4}

The transformed Stiefel problem in Eq.~\eqref{eq:stiefel_obj_v4} has $n_{\mathrm{free}}\gg 1$ for the large-scale regime, and the Stiefel iterate has $r=2$ columns corresponding to the $x$- and $y$-coordinates.
We solve it by SSM on the Stiefel manifold, using sparse operator evaluations in the ambient space and small dense solves in an adaptively built low-dimensional subspace.

\paragraph{SSM iteration}
Large-scale Stiefel problems with $n$ very large but $r$ small are well suited to SSMs, which alternate between
(i) building a low-dimensional subspace that captures locally important directions, and
(ii) solving a reduced Stiefel problem in that subspace.
A typical iteration constructs an orthonormal basis $\bm V_k\in\mathbb{R}^{n_{\mathrm{free}}\times d_k}$, where $d_k\ll n_{\mathrm{free}}$, and solves $\min_{\bm{\check X}\in \mathrm{St}(d_k,r)} \hat f_k(\bm{\check X})$, then sets $\bm{\tilde X}_{k+1}=\bm V_k \bm{\check X}$.
The subspace is enriched with the current iterate, the Euclidean gradient direction $\widehat{\bm L}\bm{\tilde X}_k\bm C+\bm B$, curvature-informed directions (Newton-CG step), and low-frequency eigenvectors of the quadratic operator~\cite{ChenCH24_StiefelSSM_arXiv}.

\paragraph{Quadratic structure and gradients}
For symmetric matrices $\widehat{\bm L}$ and $\bm C$, the Euclidean gradient of Eq.~\eqref{eq:stiefel_obj_v4} is
\begin{equation}
\label{eq:euclid_grad_v4}
\nabla f(\bm{\tilde X})=2\,(\widehat{\bm L}\, \bm{\tilde X} \bm C + \bm B).
\end{equation}
The Riemannian gradient and first-order
conditions then follow from
Eq.~\eqref{eq:riem_grad_prelim}--\eqref{eq:polar_retraction}.

\paragraph{Ground subspace and spectral shift~\cite{ChenCH24_StiefelSSM_arXiv}}
A key numerical issue in large graph-based Stiefel problems is the presence of very small nonzero eigenvalues in $\widehat{\bm L}$, which correspond to low-frequency graph
modes. 
Since the projector $\bm P$ removes the trivial mode, we compute a small ``ground'' subspace $\bm V_g=[\bm v_1,\ldots,\bm v_r]$ of the smallest nontrivial eigenvectors of $\widehat{\bm L}$ with eigenvalues $0<d_1\le\cdots\le d_r$.
We then define a spectrally shifted operator
\begin{equation}
\label{eq:spectral_shift_v4}
\widehat{\bm L}_e
=
\widehat{\bm L} + \bm V_g\big(d_r \bm I - \mathrm{diag}(d_1,\ldots,d_r)\big)\bm V_g^\top,
\end{equation}
which lifts the smallest eigenvalues within $\mathrm{span}(\bm V_g)$ to match the $r$-th eigenvalue $d_r$, while leaving the operator unchanged outside that subspace.
This construction improves conditioning without destroying sparsity, since $\widehat{\bm L}_e$ differs from $\widehat{\bm L}$ by a rank-at-most-$r$ update. The rank is exactly $r-1$ when $d_1<\cdots<d_{r-1}<d_r$.
We follow the formulation of~\cite{ChenCH24_StiefelSSM_arXiv}, which targets the indefiniteness inside $\mathrm{span}(\bm V_g)$. This is more selective than the global multiplier-dependent identity shift used in the earlier sphere-constrained formulation~\cite{ChenCCHLW23_ISPD}, which adds a constant-magnitude shift to the entire operator.

At iterate $\bm{\tilde X}_k$, we form the consistent linear term $\bm B_k$ from the matched-stationarity condition
\begin{equation}
\label{eq:Bk_v4}
\begin{aligned}
\widehat{\bm L}_e\,\bm{\tilde X}_k\,\bm C + \bm B_k = \widehat{\bm L}\,\bm{\tilde X}_k\,\bm C + \bm B,
\\\text{i.e.,}\quad
\bm B_k = \bm B - (\widehat{\bm L}_e-\widehat{\bm L})\,\bm{\tilde X}_k\,\bm C,
\end{aligned}
\end{equation}
so that the surrogate based on $\widehat{\bm L}_e$ has the same gradient as the original objective at $\bm{\tilde X}_k$.

\paragraph{Subspace formation for reduced solve}
Each outer iteration first computes a tangent-space direction
by a safeguarded Newton-CG solve. The safeguarding combines
two ingredients of~\cite{ChenCH24_StiefelSSM_arXiv}: (i) the
spectral shift $\widehat{\bm L}_e$ that lifts the ground eigenvalues
to $d_r$, and (ii) a multiplier safeguard that clips the
generalized eigenvalues of the Stiefel multiplier.

At outer iteration $k$, let
\[
\sigma_k =
\sigma_{\min}(\bm V_g^{\!\top}\bm B_k\bm C^{-1}).
\]
For a symmetric multiplier $\bm\Lambda$, define the safeguard
map $s_k(\bm\Lambda)$ by the eigendecomposition
\[
\bm C^{-1/2}\bm\Lambda\bm C^{-1/2}
=
\bm U\operatorname{diag}(\gamma_1,\ldots,\gamma_r)\bm U^\top
\]
and
\[
s_k(\bm \Lambda)
=
\bm C^{1/2}\bm U
\operatorname{diag}
\left(\min\{\gamma_i,d_r-\sigma_k\}\right)_{i=1}^r
\bm U^\top \bm C^{1/2}.
\]
By construction,
\[
s_k(\bm \Lambda) \preceq (d_r-\sigma_k)\bm C \preceq d_r\bm C.
\]

The Newton-CG operator uses $s_k(\bm \Lambda)$ in the multiplier
term. Together with the constraint $\bm V_g^\top \bm Z_k = \bm 0$ on the
Newton step, this safeguard keeps the projected Newton system
positive semidefinite and suitable for conjugate-gradient
solution.

We then form a small basis $\bm V_k\in\mathbb{R}^{n_{\mathrm{free}}\times d_k}$ by orthonormalizing the columns of
\begin{equation}
\label{eq:ssm_subspace_v4}
\bm{\mathcal S}_k=[\,\bm{\tilde X}_k,\ \widehat{\bm L}\,\bm{\tilde X}_k\bm C+\bm B,\ \bm V_g,\ \bm Z_k\,],
\end{equation}
following~\cite{ChenCH24_StiefelSSM_arXiv}: the current iterate, the Euclidean gradient direction $\widehat{\bm L}\,\bm{\tilde X}_k\bm C+\bm B$, the ground eigenvectors, and the Newton direction, each contributing $r$ columns.
We use the Euclidean rather than the Riemannian gradient here because adding the projection $-\bm{\tilde X}_k(\bm{\tilde X}_k^{\!\top}\nabla f(\bm{\tilde X}_k))_{\mathrm{sym}}$ would only contribute a column already in $\mathrm{span}(\bm{\tilde X}_k)$, leaving the subspace unchanged.
The resulting subspace dimension is $d_k\le 4r$. With $r=2$ this gives $d_k\le 8$, so the reduced solve in Eq.~\eqref{eq:reduced_v4} operates on matrices of size at most $8\times 2$.
We then solve the reduced Stiefel problem
\begin{equation}
\begin{gathered}
\label{eq:reduced_v4}
\min_{\bm{\check X}\in\mathrm{St}(d_k,2)}
\ \
\hat f_k(\bm{\check X}),\\
\hat f_k(\bm{\check X})
=
\mathrm{tr}\!\big(\bm{\check X}^\top (\bm V_k^\top \widehat{\bm L}_e \bm V_k)\,\bm{\check X}\,\bm C\big)
+2\,\mathrm{tr}\!\big((\bm V_k^\top \bm B_k)^\top \bm{\check X}\big),
\end{gathered}
\end{equation}
using a small number of Riemannian Newton steps with polar retraction (Eq.~\eqref{eq:polar_retraction}), and update the full iterate by
\begin{equation}
\label{eq:ssm_update_v4}
\bm{\tilde X}_{k+1}=\bm V_k \bm{\check X}.
\end{equation}
\paragraph{Algorithm summary}
The full SSM iteration is summarized in Algorithm~\ref{alg:ssm_v4}, and the inner Riemannian Newton solver used in the reduced subspace is detailed in Algorithm~\ref{alg:inner_newton_v4}.
The pre-computation of $(\widehat{\bm L},\bm B,\bm C)$ from the netlist hypergraph follows the construction in Sections~\ref{subsec:qcqp_formulation}--\ref{subsec:stiefel_transform_solver}: the sparse Laplacian $\bm L$ is partitioned into $(\bm L_{FF},\bm L_{FM},\bm L_{MF},\bm L_{MM})$ and the spreading targets $(c_1,c_2,c_3)$ are computed via Eq.~\eqref{eq:moment_constants_v4}. The linear term $\bm E_0$ (Eq.~\eqref{eq:E0_v4}) and constraint matrix $\bm C$ in Eq.~\eqref{eq:C_construct_v4} are then formed, $\bm C$ is rescaled by $\rho^2$ in Eq.~\eqref{eq:C_scale_v4_eq}, and metric equalization in Eq.~\eqref{eq:equalization_v4} together with the rank-one projector $\bm P$ in Eq.~\eqref{eq:P_v4} yields $\widehat{\bm L}=\bm P\widetilde{\bm L}\bm P$ and $\bm B=\bm P\widetilde{\bm E}_0$ in Eq.~\eqref{eq:AB_v4}.


Algorithm~\ref{alg:ssm_v4} implements the Stiefel-SSM
iteration with periodic Laplacian updates. Lines~1--7 perform
a one-time setup: lines~1--2 compute the ground subspace
$\bm V_g$, lines~3--4 form the spectrally shifted operator
$\widehat{\bm L}_e$, and lines~5--7 initialize the Stiefel
iterate via singular value decomposition (SVD) of
$\bm V_g^{\!\top}\bm B$. The outer loop (lines~8--27) runs
until the Riemannian-gradient norm falls below $\varepsilon$.
Lines~10--11, executed every $T_{\mathrm{upd}}$ iterations,
periodically rebuild the reduced graph using the multi-pin
sparsification of Section~\ref{subsec:sparsification} and
reassemble the operator pieces. Each iteration then updates the
consistent linear term $\bm B_k$ matched to the spectral shift
(lines~13--14), computes the safeguard value $\sigma_k$ and a
Newton-CG tangent direction $\bm Z_k$ orthogonal to the ground
subspace (lines~15--17), assembles the SSM subspace basis
$\bm V_k$ (lines~18--19), forms the reduced operator and
linear term
$(\bar{\bm L}_k,\bar{\bm B}_k)
=
(\bm V_k^{\!\top}\widehat{\bm L}_e\bm V_k,\,
 \bm V_k^{\!\top}\bm B_k)$
with initial point
$\bm Y_{1,k}=\bm V_k^{\!\top}\bm{\tilde X}_k$
(line~21), and invokes Algorithm~\ref{alg:inner_newton_v4}
with
$(\bar{\bm L}_k,\bar{\bm B}_k,\bm C,\bm Y_{1,k},
d_r,\sigma_k,\varepsilon_{\rm in},j_{\max})$
on the reduced subspace (line~22). The full Stiefel iterate is
updated in line~23. Lines~28--30 map the converged Stiefel
iterate back to physical coordinates.

\begin{algorithm}[h!]
\caption{Stiefel-SSM solver}
\label{alg:ssm_v4}
\begin{algorithmic}[1]
\REQUIRE Projected equalized operator $\widehat{\bm L}$,
projected linear term $\bm B$, constraint matrix $\bm C$;
update period $T_{\mathrm{upd}}$;
outer tolerance $\varepsilon$, inner tolerance $\varepsilon_{\mathrm{in}}$,
maximum inner Newton iterations $j_{\max}$;
back-mapping data $\bm G_M,\Delta_x,\Delta_y$
\ENSURE $\bm{\tilde X}\in\mathrm{St}(n_{\mathrm{free}},2)$, $\bm X_M$
\STATE \textit{// Ground subspace}
\STATE $(\bm V_g,d_1,\ldots,d_r)\leftarrow$ ground-eigenpairs of $\widehat{\bm L}$
\STATE \textit{// Spectral shift, Eq.~\eqref{eq:spectral_shift_v4}}
\STATE $\widehat{\bm L}_e\leftarrow \widehat{\bm L} + \bm V_g(d_r\bm I-\mathrm{diag}(d_1,\ldots,d_r))\bm V_g^{\!\top}$
\STATE \textit{// Initialize Stiefel iterate via SVD}
\STATE $\bm U_E\bm\Sigma\bm V_E^{\!\top}\leftarrow\mathrm{svd}(\bm V_g^{\!\top}\bm B)$
\STATE $\bm{\tilde X}_0\leftarrow -\bm V_g\bm U_E\bm V_E^{\!\top}$
\FOR{$k=0,1,2,\ldots$}
    \IF{$k>0$ \textbf{and} $k\bmod T_{\mathrm{upd}}=0$}
        \STATE \textit{// Periodic Laplacian update (Section~\ref{subsec:sparsification})}
        \STATE reassemble $(\widehat{\bm L},\bm B,\bm V_g,d_r,\widehat{\bm L}_e)$ from $\bm{\tilde X}_k$
    \ENDIF
    \STATE \textit{// Matched linear term, Eq.~\eqref{eq:Bk_v4}}
    \STATE $\bm B_k\leftarrow \bm B-(\widehat{\bm L}_e-\widehat{\bm L})\bm{\tilde X}_k\bm C$
    \STATE \textit{// Tangent direction via Newton-CG}
    \STATE $\sigma_k \leftarrow \sigma_{\min}(\bm V_g^\top \bm B_k \bm C^{-1})$
    \STATE $\bm Z_k\leftarrow$ Newton-CG$(\widehat{\bm L}_e,\bm B_k,\bm C,\bm{\tilde X}_k, d_r, \sigma_k)$ \\ with $\bm V_g^{\!\top}\bm Z_k=\bm 0$
    \STATE \textit{// SSM subspace, Eq.~\eqref{eq:ssm_subspace_v4}}
    \STATE $\bm V_k\leftarrow\mathrm{qr}([\bm{\tilde X}_k,\,\widehat{\bm L}\bm{\tilde X}_k\bm C+\bm B,\,\bm V_g,\,\bm Z_k])$
    \STATE \textit{// Inner Newton + update full iterate}
    \STATE $\bar{\bm L}_k\leftarrow \bm V_k^{\!\top}\widehat{\bm L}_e\bm V_k,\quad
\bar{\bm B}_k\leftarrow \bm V_k^{\!\top}\bm B_k,\quad
\bm Y_{1,k}\leftarrow \bm V_k^{\!\top}\bm{\tilde X}_k$

\STATE $\bm{\check X}\leftarrow$
Algorithm~\ref{alg:inner_newton_v4}$(\bar{\bm L}_k,\bar{\bm B}_k,\bm C,\bm Y_{1,k},d_r,\sigma_k,\varepsilon_{\mathrm{in}},j_{\max})$

\STATE $\bm{\tilde X}_{k+1}\leftarrow \bm V_k\bm{\check X}$
    \IF{$\|\mathrm{grad}\,f(\bm{\tilde X}_{k+1})\|_F\le\varepsilon$}
        \STATE \textbf{break}
    \ENDIF
\ENDFOR
\STATE \textit{// Back-mapping to physical coordinates}
\STATE $\bm{\tilde X}_M\leftarrow \bm G_M^{-1/2}\bm{\tilde X}_{k+1}\bm C^{1/2}$
\STATE $\bm X_M\leftarrow \bm{\tilde X}_M+\mathbf{1}[\Delta_x\ \Delta_y]$
\end{algorithmic}
\end{algorithm}


Algorithm~\ref{alg:inner_newton_v4} implements the inner
Riemannian Newton solve on the reduced subproblem
(Eq.~\eqref{eq:reduced_v4}). Lines~1--2 define the safeguard
map $s_k(\cdot)$ and initialize the raw symmetric Lagrange
multiplier $\bm\Lambda_1$; the raw multiplier itself is not
clipped. Each Newton iteration (lines~3--15) computes the
stationarity residual $\bm E_j$ using the raw multiplier
$\bm\Lambda_j$ (lines~4--5), solves the projected Newton
system for the tangent step $\bm Z_j$ using $s_k(\bm\Lambda_j)$
in the multiplier term (lines~6--7), takes a line-searched
polar-retraction step to obtain $\bm Y_{j+1}$ (lines~8--9),
updates the raw symmetric multiplier at the new point
(lines~10--11), and checks convergence (lines~12--14). The
final reduced iterate $\bm{\check X}$ is returned in line~16.
This separation keeps the residual tied to the current
symmetric multiplier, while the Newton-CG operator uses a
safeguarded multiplier satisfying
$s_k(\bm\Lambda_j)\preceq d_r\bm C$.

\begin{algorithm}[h!]
\caption{Inner Riemannian Newton}
\label{alg:inner_newton_v4}
\begin{algorithmic}[1]
\REQUIRE Reduced operator $\bar{\bm L}\in\mathbb R^{d_k\times d_k}$,
reduced linear term $\bar{\bm B}\in\mathbb R^{d_k\times 2}$,
constraint matrix $\bm C\in\mathbb R^{2\times 2}$,
initial point $\bm Y_1\in\mathrm{St}(d_k,2)$,
ground eigenvalue $d_r$, safeguard value $\sigma_k$,
tolerance $\varepsilon_{\mathrm{in}}$, maximum iterations $j_{\max}$
\ENSURE $\bm{\check X}\in\mathrm{St}(d_k,2)$
\STATE \textit{// Define $s_k(\bm\Lambda)$ by clipping eigenvalues of
   $\bm C^{-1/2}\bm\Lambda \bm C^{-1/2}$ at $d_r-\sigma_k$}
\STATE $\bm\Lambda_1\leftarrow(\bm Y_1^{\!\top}(\bar{\bm L}\bm Y_1\bm C+\bar{\bm B}))_{\mathrm{sym}}$
\FOR{$j=1,2,\ldots,j_{\max}$}
    \STATE \textit{// Newton residual}
    \STATE $\bm E_j\leftarrow -\bar{\bm L}\bm Y_j\bm C-\bar{\bm B}+\bm Y_j\bm\Lambda_j$
    \STATE \textit{// Projected Newton step, solve for $\bm Z_j$}
    \STATE Solve $\mathrm{Proj}_{\bm Y_j}\!\{\bar{\bm L}(\mathrm{Proj}_{\bm Y_j}\bm Z_j)\bm C-(\mathrm{Proj}_{\bm Y_j}\bm Z_j)\,s_k(\bm\Lambda_j)\}$ \\
    \hphantom{\textbf{Solve}}~$=\mathrm{Proj}_{\bm Y_j}\bm E_j$
    \STATE \textit{// Polar retraction, Eq.~\eqref{eq:polar_retraction}}
    \STATE $\alpha\leftarrow$ line-search,\quad $\bm Y_{j+1}\leftarrow \mathrm{Retr}_{\bm Y_j}(\alpha\bm Z_j)$
    \STATE \textit{// Update $\bm\Lambda$}
    \STATE $\bm\Lambda_{j+1}\leftarrow (\bm Y_{j+1}^{\!\top}(\bar{\bm L}\bm Y_{j+1}\bm C+\bar{\bm B}))_{\mathrm{sym}}$
    \IF{$\|\bm E_j\|_F\le\varepsilon_{\mathrm{in}}$}
        \STATE \textbf{break}
    \ENDIF
\ENDFOR
\STATE $\bm{\check X}\leftarrow \bm Y_{j+1}$
\end{algorithmic}
\end{algorithm}


\subsection{Multi-Pin Geometric Sparsification}
\label{subsec:sparsification}

This subsection describes how the sparse Laplacian $\bm L$ is constructed and periodically updated from multi-pin nets.

The Laplacian $\bm L$ is obtained by reducing multi-pin nets to sparse graph edges.
To keep $\bm L$ sparse while reflecting current geometry, we use an \emph{angle-sorted cycle} model for each multi-pin net and rebuild the graph periodically.

\begin{figure}[!htbp]
\centering
\begin{tikzpicture}[
  pin/.style={circle, fill=blue!60!black, inner sep=0pt,
    minimum size=4pt},
  cent/.style={circle, draw=red!60, fill=red!20, inner sep=0pt,
    minimum size=5pt},
  edge/.style={draw=blue!40, thick},
  cyc/.style={draw=blue!60!black, thick},
  ang/.style={draw=darkgray!100, dashed, thin},
  lbl/.style={font=\footnotesize\sffamily},
]

\node[lbl] at (0, 1.7) {\textbf{(a)} Multi-pin net};

\node[cent, label={[lbl,red!60]below:$(\bar x,\bar y)$}]
  (C) at (0, 0) {};

\node[pin, label={[lbl]right:$p_1$}] (p1) at ( 1.1,  0.4) {};
\node[pin, label={[lbl]above:$p_2$}] (p2) at ( 0.3,  1.0) {};
\node[pin, label={[lbl]left:$p_3$}]  (p3) at (-0.9,  0.6) {};
\node[pin, label={[lbl]left:$p_4$}]  (p4) at (-1.0, -0.3) {};
\node[pin, label={[lbl]below:$p_5$}] (p5) at (-0.2, -0.9) {};
\node[pin, label={[lbl]right:$p_6$}] (p6) at ( 0.8, -0.7) {};

\foreach \p in {p1,p2,p3,p4,p5,p6} {
  \draw[ang] (C) -- (\p);
}

\node[lbl, text=darkgray] at (0.7, 0.1) {$\theta_1$};
\node[lbl, text=darkgray] at (0.15, 0.55) {$\theta_2$};

\begin{scope}[xshift=4.0cm]
\node[lbl] at (0, 1.7) {\textbf{(b)} Angle-sorted cycle};

\node[pin] (q1) at ( 1.1,  0.4) {};
\node[pin] (q2) at ( 0.3,  1.0) {};
\node[pin] (q3) at (-0.9,  0.6) {};
\node[pin] (q4) at (-1.0, -0.3) {};
\node[pin] (q5) at (-0.2, -0.9) {};
\node[pin] (q6) at ( 0.8, -0.7) {};

\draw[cyc] (q1) -- (q2);
\draw[cyc] (q2) -- (q3);
\draw[cyc] (q3) -- (q4);
\draw[cyc] (q4) -- (q5);
\draw[cyc] (q5) -- (q6);
\draw[cyc] (q6) -- (q1);

\node[lbl, below=0.3cm of q5] {$m$ pins $\to$ $m$ edges};
\end{scope}

\end{tikzpicture}
\caption{Angle-sorted cycle sparsification for a
  multi-pin net.
  (a)~Compute centroid $(\bar x, \bar y)$ and
  angles $\theta_i$ from centroid to each pin.
  (b)~Sort pins by angle and form a cycle,
  producing $m$ edges (vs.\ $m(m{-}1)/2$ for the
  clique model).
  The cycle is rebuilt every $T_{\mathrm{upd}}$
  iterations to reflect updated geometry.}
\label{fig:cycle_sparsification}
\end{figure}
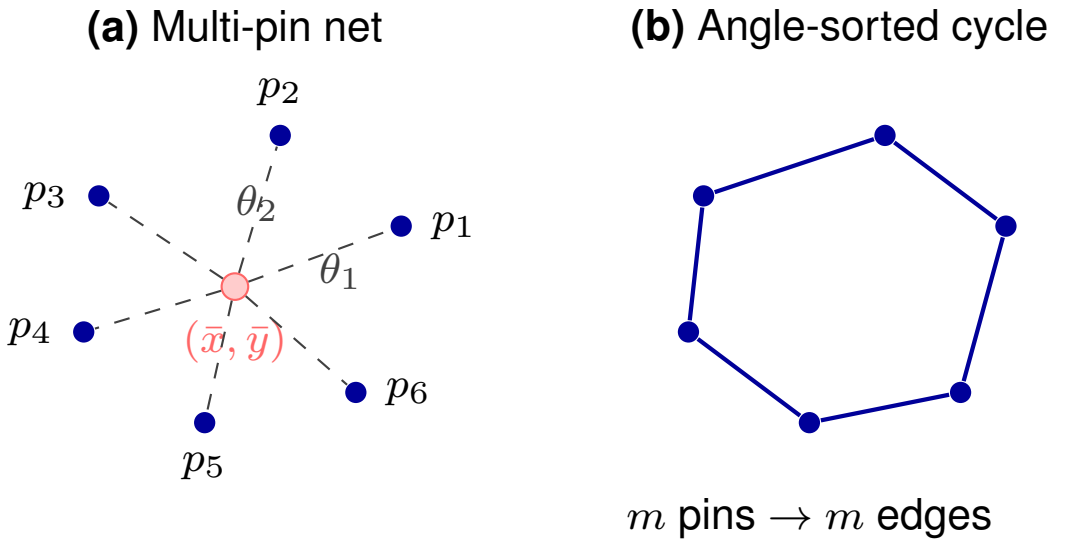

For a multi-pin net with pin locations $\{(x_i,y_i)\}_{i=1}^{m}$, compute its centroid
\[
(\bar x,\bar y)=\Big(\frac{1}{m}\sum_{i=1}^{m}x_i,\ \frac{1}{m}\sum_{i=1}^{m}y_i\Big),
\]
compute angles $\theta_i=\mathrm{atan2}(y_i-\bar y,\ x_i-\bar x)$, sort pins by $\theta_i$, and connect the sorted pins into a cycle.
Each two-pin net is a unit-weight edge, and each cycle edge has weight $1/2$.
This uses $O(m)$ edges per net, compared to $O(m^2)$ for the clique model and $O(m)$ for the star model but without adding a star-center node~\cite{ChenCCHLW23_ISPD}.
The cycle model captures local geometric proximity through the cyclic neighborhood and naturally adapts to the current pin configuration when the graph is rebuilt.
Fig.~\ref{fig:cycle_sparsification} illustrates
the construction.

Because the cycle depends on current coordinates, we couple it with an update loop.
Every $T_{\mathrm{upd}}$ outer iterations, we rebuild the reduced graph, update $\bm L$, and reassemble $(\bm L_{MM},\bm L_{MF})$, $\bm E$, $\bm E_0$, and the projected operator pieces $(\widehat{\bm L},\bm B)$.
Between updates, we optimize Eq.~\eqref{eq:stiefel_obj_v4} using the Stiefel-SSM iterations above.
This loop differs from the iterative reweighting of~\cite{ChenCCHLW23_ISPD}, which adjusts edge weights as a function of current $\ell_2$ distance to drive the squared-wirelength approximation toward $\ell_1$ HPWL minimization. Our update instead rebuilds the cycle topology by angle order while keeping uniform edge weights, targeting the squared-wirelength approximation without explicit $\ell_1$ relaxation.


\section{Experimental Results}
\label{sec:experiments}

This section evaluates the proposed Stiefel-constrained
initialization as a drop-in initializer for three
state-of-the-art mixed-size placers:
~\emph{RePlAce}~\cite{ChengKKW19_RePlAce},
~\emph{DREAMPlace~4.1}~\cite{LinDLRKP19_DREAMPlace, LiaoEtAl_StrongerMixedSize},
and ~\emph{AutoDMP}~\cite{AutoDMP23_ISPD}.
We report pre-route placement quality on 20 benchmarks
(Section~\ref{subsec:preroute}) and post-route
physical-design quality on three benchmarks
(Section~\ref{subsec:postroute}).

\subsection{Experimental Setup}
\label{subsec:exp_setup}

\subsubsection{Benchmarks}
Tables~\ref{tab:preroute_tilos} and~\ref{tab:preroute_mms} summarize the benchmark
suite and pre-route results.
We adopt the same benchmark collection
as~\cite{LiaoEtAl_StrongerMixedSize}: the 16 MMS benchmarks~\cite{YanViswanathanChu09_DAC_MixedSizeComplexities} (as bundled with DREAMPlace~\cite{LinDLRKP19_DREAMPlace}) and the four TILOS MacroPlacement benchmarks~\cite{TILOS_MacroPlacement}.
The TILOS repository provides netlists for Ariane133 (ASAP7~\cite{ClarkEtAl16_ASAP7}), Mempool tile (ASAP7), NVDLA (NanGate45~\cite{Nangate45_OCL}), and BlackParrot (NanGate45).

For the post-route study (Section~\ref{subsec:postroute}) we use a
separate benchmark set from the updated assessment by Cheng et al.~\cite{KahngEtAl_UpdatedAssessment}:
Ariane133, BlackParrot, and Mempool group, all targeting the ASAP7 technology node.
Mempool tile and Mempool group are distinct benchmark configurations: Mempool tile is used in the pre-route TILOS evaluation, whereas Mempool group is used only in the post-route study from ~\cite{KahngEtAl_UpdatedAssessment}.

I/O pins are treated as fixed objects in $\mathcal{F}$
(Section~\ref{subsec:prelim_mixedsize}).
The benchmark suite spans a range of I/O configurations:
some benchmarks place I/O pins on the die boundary,
others outside the boundary, and others in the interior.
Fig.~\ref{fig:placement_ariane}--\ref{fig:placement_newblue3}
show representative examples.
The proposed formulation handles all such cases uniformly: coordinates of fixed objects enter the reduced objective through their net connections to movable objects, while fixed-object subtraction and centering correction are incorporated into the movable-variable QCQP constraints (Section~\ref{subsec:problem_construction}). Thus, boundary, exterior, and interior I/O pins require no case-specific handling.

\FloatBarrier
\begin{table*}[!htbp]
\centering
\caption{Benchmark characteristics and pre-route wHPWL ($\times 10^{6}$) on the four TILOS benchmarks~\cite{TILOS_MacroPlacement}, lower is better.\\
  ``w/o''~/~``w/'' denote results without / with the proposed initialization.
  \textbf{Impr.\,(\%)} is the wHPWL reduction.\\
  \textbf{Init (min)} is the wall-clock runtime of the initializer.
  Best per benchmark is in \textbf{bold}.}
\label{tab:preroute_tilos}
\vspace{-6pt}
\begin{footnotesize}
\setlength{\tabcolsep}{5.5pt}
\begin{tabular}{l rrr rrr rrr rrr r}
\toprule
 & & &
 & \multicolumn{3}{c}{RePlAce~\cite{ChengKKW19_RePlAce}}
 & \multicolumn{3}{c}{DREAMPlace~\cite{LinDLRKP19_DREAMPlace, LiaoEtAl_StrongerMixedSize}}
 & \multicolumn{3}{c}{AutoDMP~\cite{AutoDMP23_ISPD}}
 & Init \\
\cmidrule(lr){5-7}
\cmidrule(lr){8-10}
\cmidrule(lr){11-13}
Benchmark
 & Cells & Macros & Nets
 & w/o & w/ & Impr.
 & w/o & w/ & Impr.
 & w/o & w/ & Impr.
 & (min) \\
\midrule
Ariane133
 & 98K & 133 & 101K
 & 10.46 & 9.31 & 11.0
 & 9.84 & 9.04 & 8.1
 & 9.11 & \textbf{8.60} & 5.6
 & 2.4 \\
Mempool tile
 & 132K & 20 & 135K
 & 11.19 & \textbf{10.52} & 6.0
 & 10.96 & 10.63 & 3.0
 & 11.05 & 10.67 & 3.4
 & 3.2 \\
NVDLA
 & 153K & 128 & 208K
 & 102.31 & 65.25 & 36.2
 & 36.24 & \textbf{34.09} & 6.0
 & 38.29 & 34.50 & 9.9
 & 3.8 \\
BlackParrot
 & 1.28M & 220 & 1.5M
 & 158.73 & 155.42 & 2.1
 & 143.36 & 130.00 & 9.3
 & 130.93 & \textbf{122.90} & 6.1
 & 29.6 \\
\midrule
\textbf{Avg.}
 & & & &  &  & \textbf{13.8}
 &  &  & \textbf{6.6}
 &  &  & \textbf{6.3}
 & \textbf{9.8} \\
\bottomrule
\end{tabular}
\end{footnotesize}
\end{table*}

\FloatBarrier
\begin{table*}[!htbp]
\centering
\caption{Benchmark characteristics and pre-route wHPWL ($\times 10^{6}$) on the 16 MMS benchmarks~\cite{YanViswanathanChu09_DAC_MixedSizeComplexities}, which are bundled with DREAMPlace~\cite{LinDLRKP19_DREAMPlace, LiaoEtAl_StrongerMixedSize}, lower is better.
  ``w/o''~/~``w/'' denote results without / with the proposed initialization.
  \textbf{Impr.\,(\%)} is the wHPWL reduction.
  \textbf{Init (min)} is the wall-clock runtime of the initializer.
  Best per benchmark in \textbf{bold}.
  Designs marked N/C did not converge and are excluded from averages.
  RePlAce is omitted because its mixed-size flow does not accept the MMS benchmark format.}
\label{tab:preroute_mms}
\vspace{-6pt}
\begin{footnotesize}
\setlength{\tabcolsep}{5.5pt}
\begin{tabular}{l rrr rrr rrr r}
\toprule
 & & &
 & \multicolumn{3}{c}{DREAMPlace~\cite{LinDLRKP19_DREAMPlace, LiaoEtAl_StrongerMixedSize}}
 & \multicolumn{3}{c}{AutoDMP~\cite{AutoDMP23_ISPD}}
 & Init \\
\cmidrule(lr){5-7}
\cmidrule(lr){8-10}
Benchmark
 & Cells & Macros & Nets
 & w/o & w/ & Impr.
 & w/o & w/ & Impr.
 & (min) \\
\midrule
adaptec1
 & 211K & 63 & 221K
 & 65.22 & \textbf{63.86} & 2.1
 & 70.26 & 64.93 & 7.6
 & 5.1 \\
adaptec2
 & 254K & 127 & 266K
 & 75.12 & 71.18 & 5.2
 & 75.26 & \textbf{71.12} & 5.5
 & 6.5 \\
adaptec3
 & 451K & 58 & 467K
 & 152.70 & \textbf{152.20} & 0.3
 & N/C & 152.27 & N/C
 & 12.4 \\
adaptec4
 & 495K & 69 & 516K
 & 141.56 & \textbf{140.08} & 1.0
 & 143.40 & 141.43 & 1.4
 & 13.3 \\
adaptec5
 & 842K & 76 & 868K
 & 309.24 & 301.20 & 2.6
 & 261.10 & \textbf{248.85} & 4.7
 & 24.8 \\
bigblue1
 & 278K & 32 & 284K
 & 85.43 & \textbf{84.41} & 1.2
 & 90.51 & 84.85 & 6.3
 & 6.9 \\
bigblue2
 & 535K & 959 & 577K
 & 124.98 & \textbf{124.35} & 0.5
 & 127.40 & 125.32 & 1.6
 & 15.0 \\
bigblue3
 & 1.09M & 2549 & 1.12M
 & 271.86 & \textbf{263.35} & 3.1
 & 286.11 & 265.84 & 7.1
 & 35.6 \\
bigblue4
 & 2.17M & 199 & 2.23M
 & 644.87 & \textbf{641.56} & 0.5
 & 669.22 & 656.18 & 1.9
 & 87.8 \\
newblue1
 & 330K & 64 & 339K
 & 59.78 & \textbf{56.62} & 5.3
 & 58.92 & 57.72 & 2.0
 & 9.4 \\
newblue2
 & 437K & 3748 & 465K
 & 150.96 & \textbf{149.77} & 0.8
 & 154.66 & 150.32 & 2.8
 & 13.3 \\
newblue3
 & 483K & 51 & 552K
 & 297.72 & \textbf{256.07} & 14.0
 & 305.38 & 256.25 & 16.1
 & 13.4 \\
newblue4
 & 643K & 81 & 637K
 & 230.81 & 227.54 & 1.4
 & 183.35 & \textbf{179.92} & 1.9
 & 18.3 \\
newblue5
 & 1.23M & 91 & 1.28M
 & 403.12 & 390.36 & 3.2
 & 336.31 & \textbf{328.80} & 2.2
 & 39.9 \\
newblue6
 & 1.25M & 74 & 1.29M
 & 407.81 & 404.02 & 0.9
 & 401.19 & \textbf{389.35} & 3.0
 & 41.2 \\
newblue7
 & 2.48M & 161 & 2.64M
 & 883.72 & \textbf{869.51} & 1.6
 & 897.71 & 871.98 & 2.9
 & 62.8 \\
\midrule
\textbf{Avg.}
 & & & &  &  & \textbf{2.7}
 &  &  & \textbf{4.5}
 & \textbf{25.3} \\
\bottomrule
\end{tabular}
\end{footnotesize}
\end{table*}

\FloatBarrier
\begin{table}[!htbp]
\centering
\caption{Pre-route wHPWL improvement (\%) averaged by placer, summarizing Tables~\ref{tab:preroute_tilos} and~\ref{tab:preroute_mms}.
  RePlAce runs only on the four TILOS benchmarks.
  AutoDMP averages exclude adaptec3 (default AutoDMP flow does not converge): the MMS Avg is over 15 designs and the All Avg over 19,
  given in \textbf{bold}.}
\label{tab:preroute_summary}
\vspace{-6pt}
\begin{footnotesize}
\setlength{\tabcolsep}{10pt}
\begin{tabular}{l c c c}
\toprule
Placer & TILOS Avg. & MMS Avg. & All Avg. \\
\midrule
RePlAce    & 13.8 & N/A  & N/A  \\
DREAMPlace & 6.6  & 2.7  & \textbf{3.5} \\
AutoDMP    & 6.3  & 4.5  & \textbf{4.8} \\
\bottomrule
\end{tabular}
\end{footnotesize}
\end{table}

\subsubsection{Pre-Route Flow}

\paragraph{Placement engines}
We integrate the proposed initialization into three
analytical placers.
(1)~\emph{RePlAce}~\cite{ChengKKW19_RePlAce}:
an electrostatics-based global placer that supports
mixed-size instances natively.
In our flow, RePlAce supports the four TILOS benchmarks.
The MMS benchmark format used here is not accepted by 
the RePlAce mixed-size flow without conversion.
(2)~\emph{DREAMPlace~4.1}~\cite{LinDLRKP19_DREAMPlace, LiaoEtAl_StrongerMixedSize}:
In our mixed-size experiments, we use the Barzilai-Borwein (BB)-step-enabled mixed-size flow from DREAMPlace~4.1. This flow is paired with the default two-stage macro placement and legalization flow.
(3)~\emph{AutoDMP}~\cite{AutoDMP23_ISPD}:
an automated hyperparameter tuning framework
built on top of DREAMPlace.
In every experiment the proposed initializer replaces
only the default initial placement.
All subsequent global-placement and detailed-placement
stages remain unchanged.

\paragraph{Metric}
We measure pre-route quality by
wHPWL after detailed placement.
For each placer, we report the wHPWL with and without
the proposed initialization and the resulting
percentage improvement.
Positive values denote wHPWL reduction.

\subsubsection{Hardware Environment}
We run all experiments on a workstation
equipped with a 13th-generation Intel Core
i9-13900K processor (24 cores, 5.8\,GHz max turbo),
128\,GB system memory, and an NVIDIA GeForce RTX~4090 GPU. For the experiments reported in this paper, the initializer runs primarily on the CPU, with GPU acceleration limited to the computation of the ground eigenvectors \(\bm V_g\).

\subsubsection{Initialization Parameters}
The Stiefel-SSM initializer uses an outer loop of $20$ iterations
with Laplacian reconstruction period $T_{\mathrm{upd}}=10$
across all benchmarks.
The Init-time column in Tables~\ref{tab:preroute_tilos} and~\ref{tab:preroute_mms}
reports the wall-clock runtime of the initializer at these settings.

\subsubsection{Post-Route Flow}
We run the complete routing flow using
Cadence Innovus 21.10, which performs clock-tree synthesis (CTS), routing,
and routing optimization.
We select DREAMPlace as the representative open-source
placer for the post-route comparison.
RePlAce is excluded because its mixed-size flow does not
accept the MMS benchmark format and supports only the four
TILOS benchmarks.
AutoDMP adds an automated tuning loop and has one
non-convergent default run
(Tables~\ref{tab:preroute_tilos} and~\ref{tab:preroute_mms}).
Therefore, we use DREAMPlace for controlled open-source comparisons 
because, apart from the initial coordinates, the downstream 
placement and routing flow is identical with and without the proposed 
initialization. This makes the post-route comparison directly 
attributable to the initialization.
We report metrics after post-route optimization:
standard-cell area,
routed wirelength (rWL),
power,
worst negative slack (WNS),
total negative slack (TNS),
and DRC violations.
We compare two configurations:
(1)~DREAMPlace without the proposed initialization,
and (2)~DREAMPlace with the proposed initialization.

\subsection{Pre-Route Placement Quality}
\label{subsec:preroute}

Tables~\ref{tab:preroute_tilos} and~\ref{tab:preroute_mms}
report the per-design pre-route wHPWL on the four TILOS
and 16 MMS benchmarks, respectively, with and without
the proposed initialization across the three placers.
Table~\ref{tab:preroute_summary} summarizes the averages
by placer and benchmark scope.
RePlAce is averaged only over the four TILOS benchmarks
supported by its mixed-size flow.
DREAMPlace is averaged over all 20 benchmarks.
AutoDMP is averaged over the 19 benchmarks with
converged baselines. The {\it adaptec3} benchmark is reported
separately because the default AutoDMP flow does not
converge there.

\begin{figure}[!t]
  \centering
  \includegraphics[width=\columnwidth]{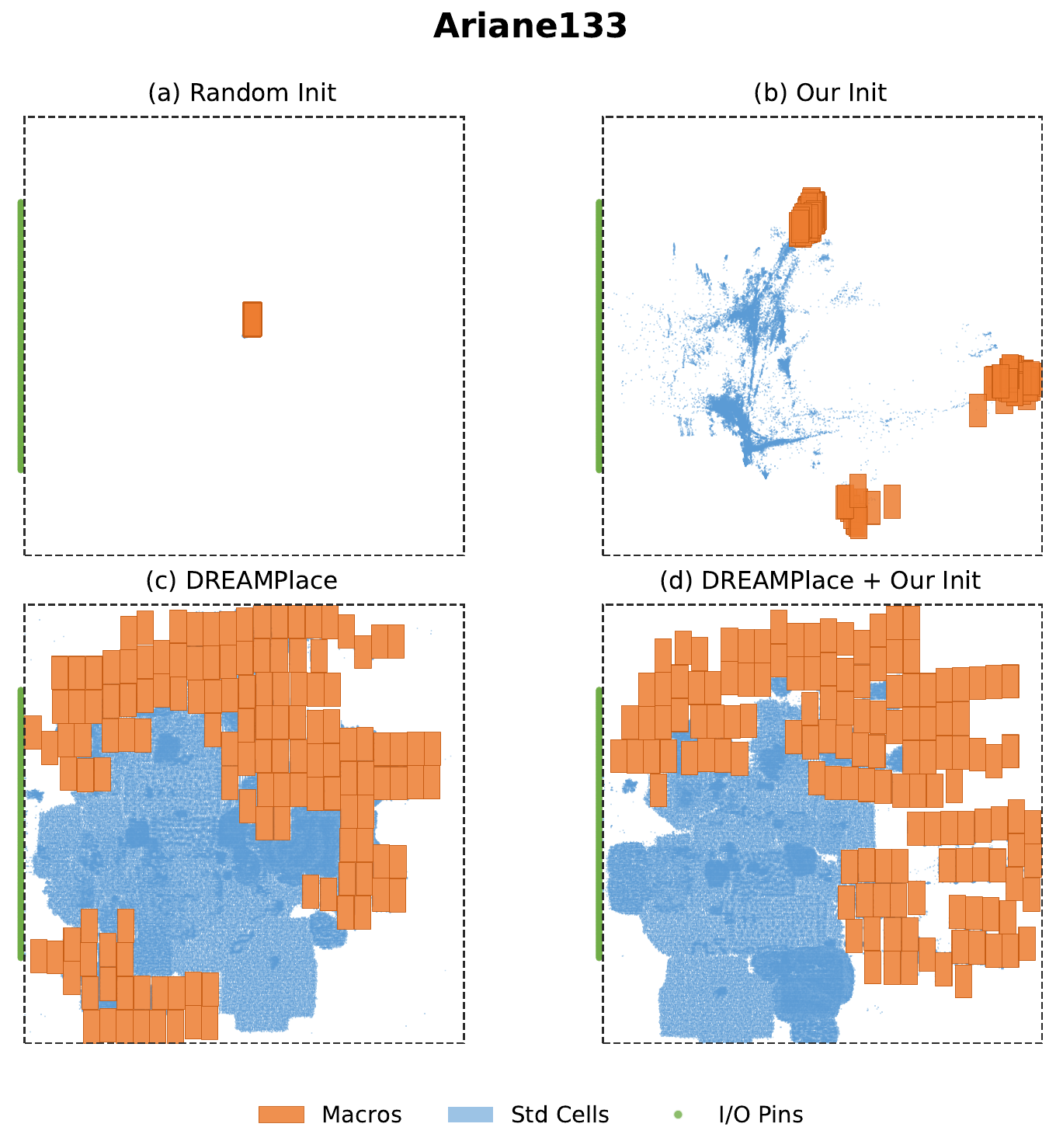}
  \caption{Macro placement for
    Ariane133 (98K cells, 133 SRAM macros).
    (a)~Random initialization,
    (b)~our initialization,
    (c)~final placement under DREAMPlace without our
    initialization,
    (d)~final placement under DREAMPlace with our
    initialization.
    Orange -- macros, blue -- standard cells,
    green -- I/O pins.}
  \label{fig:placement_ariane}
\end{figure}

The initialization consistently reduces wHPWL on every
converged run.
DREAMPlace achieves a 3.5\% average improvement  
across all 20 benchmarks, with a peak gain of 14.0\% on {\it newblue3}.
AutoDMP achieves an average improvement of 4.8\% across the
19 converged baselines, with a peak gain of 16.1\% on {\it newblue3},
and additionally converges on {\it adaptec3} when initialized
by the proposed method.
RePlAce achieves an average improvement of 13.8\% on the four
TILOS benchmarks, driven by a gain of 36.2\% on {\it NVDLA}  where the
default initialization produces highly suboptimal
macro positions.
The improvement is stable across all three
placer backends, confirming the
generality of the proposed initialization.

On BlackParrot, the largest TILOS benchmark
(1.28M cells, 220 macros), the proposed
initialization improves wHPWL across all
three placers, including a 2.1\% gain under
RePlAce and 9.3\% under DREAMPlace, confirming
that the method scales to million-cell instances.

A noteworthy case is {\it adaptec3} under AutoDMP:
the default flow fails to converge (N/C in Table~\ref{tab:preroute_mms}),
while AutoDMP with the proposed initialization converges
successfully, producing a wHPWL comparable to that of DREAMPlace.
The {\it adaptec3} result suggests that the proposed initialization 
can improve robustness for some hyperparameter-tuned flows
on challenging designs by providing a better-conditioned starting point.

Beyond the aggregate wHPWL numbers, three
representative designs reveal how the proposed
initialization improves macro arrangement
at the design level.

\subsubsection{Representative Cases}

\paragraph{Ariane133: I/O--macro separation}
Ariane133 has 133 SRAM macros and concentrates I/O
pins on the left edge.
The default DREAMPlace initialization
(Fig.~\ref{fig:placement_ariane}(c)) pulls some
macros toward the left edge, competing with
standard cells for the I/O-adjacent area.
Our initialization
(Fig.~\ref{fig:placement_ariane}(d)) pushes
macros toward the interior, freeing the
I/O-adjacent periphery for the cells that
require short wires to the pins. Wirelength
decreases by 8.1\%.

\begin{figure}[!t]
  \centering
  \includegraphics[width=\columnwidth]{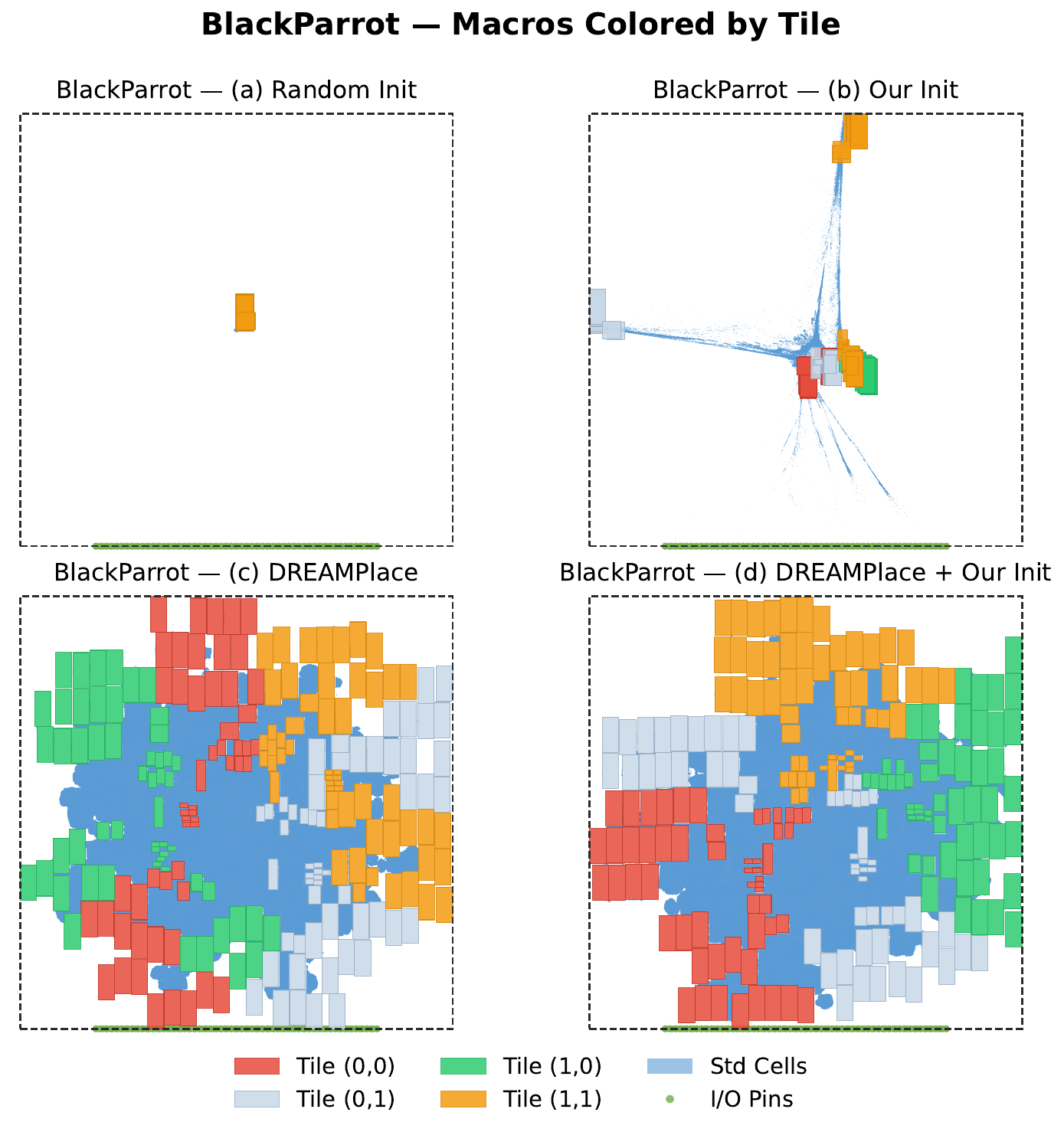}
  \caption{Macro placement for
    BlackParrot (1.28M cells, 220 macros).
    (a)~Random initialization,
    (b)~our initialization,
    (c)~final placement under DREAMPlace without our
    initialization,
    (d)~final placement under DREAMPlace with our
    initialization.
    Macros are colored by tile membership
    (Tile~(0,0)/(0,1)/(1,0)/(1,1)) in the
    $2{\times}2$ multicore layout.
    Blue -- standard cells, green -- I/O pins.}
  \label{fig:placement_blackparrot}
\end{figure}

\begin{figure}[!t]
  \centering
  \includegraphics[width=\columnwidth]{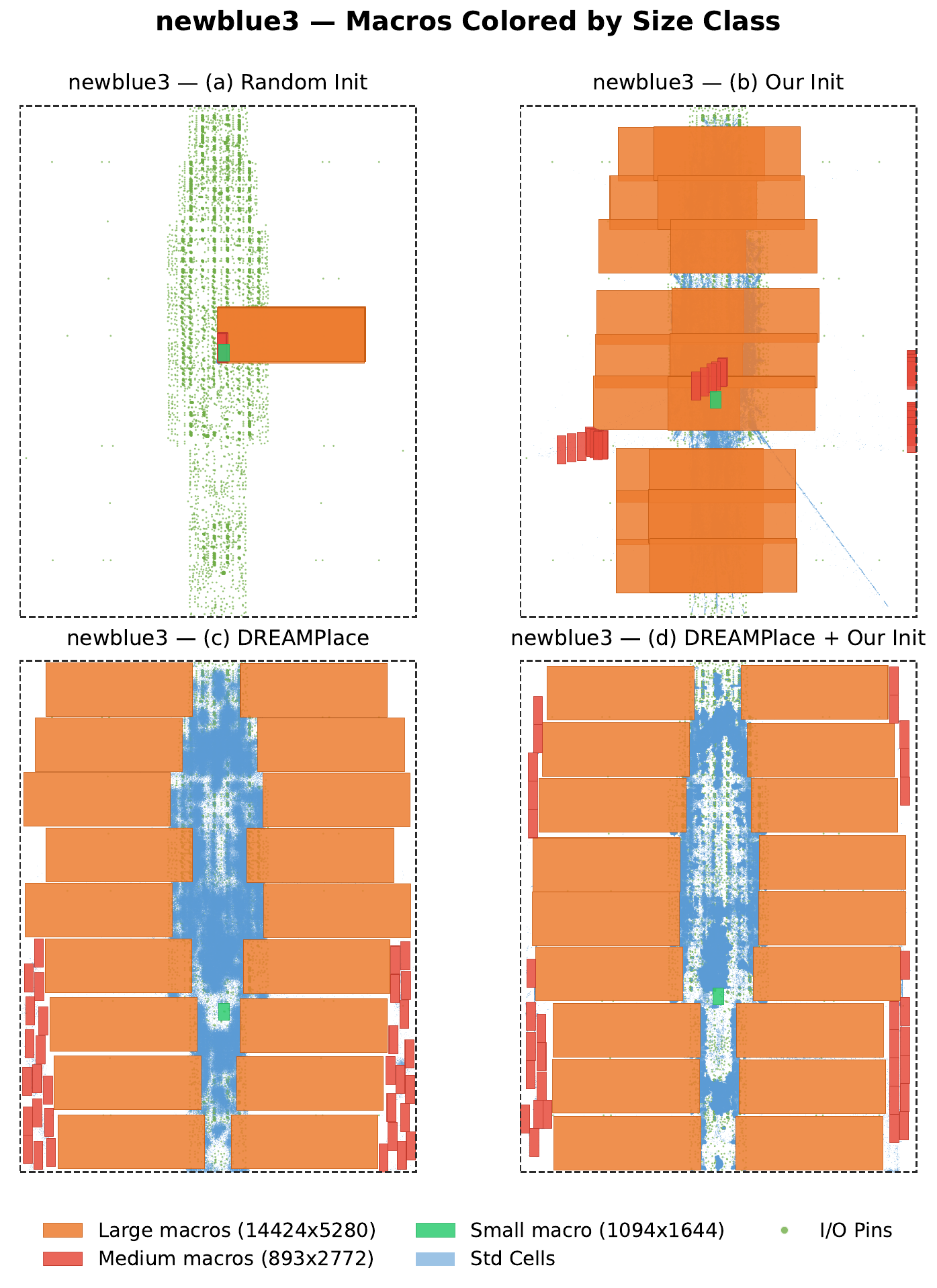}
  \caption{Macro placement for
    newblue3 (483K cells, 51 macros).
    (a)~Random initialization,
    (b)~our initialization,
    (c)~final placement under DREAMPlace without our
    initialization,
    (d)~final placement under DREAMPlace with our
    initialization.
    Macros are colored by size class (large, medium,
    small), with legend dimensions in Bookshelf grid units.
    Blue -- standard cells, green -- I/O pins.}
  \label{fig:placement_newblue3}
\end{figure}

\paragraph{BlackParrot: Tile locality}
BlackParrot (1.28M cells, 220 macros) is a
$2{\times}2$ tiled multicore processor. Each tile
has 55 macros with dense intra-tile and minimal
cross-tile connectivity.
The default initialization
(Fig.~\ref{fig:placement_blackparrot}(c))
intermixes macros from different tiles.
For example, a Tile~(0,0) DCache SRAM may end up
adjacent to a Tile~(1,1) L2 SRAM despite no
connectivity between them.
Our initialization
(Fig.~\ref{fig:placement_blackparrot}(d))
clusters macros from the same tile spatially.
This pattern is consistent with the design
intuition of keeping each tile's macros close to
its own logic and reduces inter-tile wirelength.

\paragraph{newblue3: Balanced macro distribution}
Newblue3 is dominated by 18 large horizontal macros
that confine standard cells to a narrow vertical
channel. The remaining area holds 32 medium and one
small macro.
The default initialization
(Fig.~\ref{fig:placement_newblue3}(c))
concentrates the medium macros in the lower half,
producing an asymmetric density distribution.
Our initialization
(Fig.~\ref{fig:placement_newblue3}(d))
distributes them evenly across both halves,
balancing channel density and yielding 14.0\%
wHPWL improvement under DREAMPlace and 16.1\%
under AutoDMP, the largest gains on any MMS
benchmark.

\subsubsection{Solver Convergence}
\begin{figure}[!ht]
  \centering
  \includegraphics[width=\linewidth]{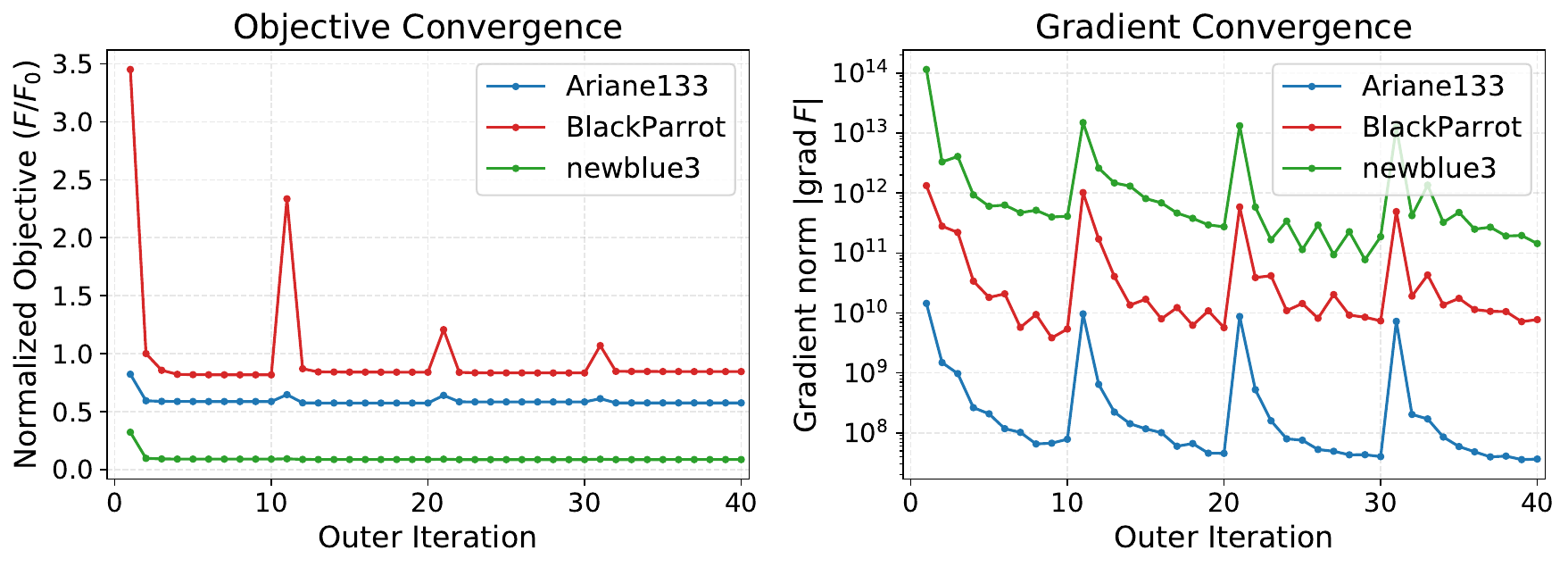}
  \caption{Convergence of Stiefel-SSM on representative
    designs.
    (Left) Normalized objective on the Stiefel manifold,
    $f(\bm{\tilde X}_k)/f(\bm{\tilde X}_0)$.
    (Right) Norm of the Riemannian gradient.
    Vertical dashed lines mark Laplacian reconstruction
    points.}
  \label{fig:convergence}
\end{figure}

Fig.~\ref{fig:convergence} shows the convergence of
Stiefel-SSM on representative designs during the
pre-route initialization stage, run for 40 outer
iterations as a diagnostic.
Tables~\ref{tab:preroute_tilos} and~\ref{tab:preroute_mms}
use the standard 20-iteration setting.
The Stiefel objective decreases within each
fixed-Laplacian interval, with controlled jumps at
the periodic Laplacian reconstruction points
(every $T_{\mathrm{upd}}=10$ outer iterations), where
the operator changes and the objective can change abruptly.
The Riemannian gradient norm trends downward
between reconstructions, consistent with the SSM
convergence guarantee~\cite{ChenCH24_StiefelSSM_arXiv}
that the gradient asymptotically vanishes
($\lim_{k\to\infty}\|\mathrm{grad}\,f(\tilde X_k)\|_F = 0$),
indicating that the placement coordinates returned
to downstream placers come from a low-gradient
initializer.

\subsection{Post-Route Physical Design Quality}
\label{subsec:postroute}

Following the post-route setup of Section~\ref{subsec:exp_setup}, Table~\ref{tab:postroute} reports post-route optimization power, performance, and area (PPA) metrics from Cadence Innovus, including remaining DRC violations.

\subsubsection{Detailed PPA Analysis}
\paragraph{Ariane133}
DREAMPlace w/ init reduces routed wirelength by 2.3\%
relative to w/o init
(959,926 vs.\ 982,089\,$\mu\mathrm{m}$),
confirming that the pre-route wHPWL gain
carries through routing.
After DRC fixing, w/ init leaves 93
remaining violations compared to 157 for w/o init,
indicating that the proposed initialization
produces a placement with fewer DRC violations.

\paragraph{BlackParrot}
On the largest benchmark (1.28M standard cells),
DREAMPlace w/ init reduces routed wirelength by
1.2\% compared to w/o init
(6,926,938 vs.\ 7,013,052\,$\mu\mathrm{m}$),
demonstrating that the initialization benefit
persists through detailed placement and routing
even at million-instance scale.

\paragraph{Mempool group}
Mempool group is the most challenging
routing scenario.
DREAMPlace w/ init reduces routed wirelength by 1.1\%
compared to w/o init
(24,233,201 vs.\ 24,499,970\,$\mu\mathrm{m}$),
with comparable power savings
(1355.7 vs.\ 1363.0\,mW).
Timing improves in TNS
($-$440.2 vs.\ $-$678.6\,ns)
while WNS is comparable
($-$0.150 vs.\ $-$0.127\,ns).
DRC violations also decrease slightly, by approximately
0.4\% (32,449 vs.\ 32,576).

\FloatBarrier
\begin{table*}[!htbp]
\centering
\caption{Post-route power, performance, and area (PPA) from Cadence Innovus on ASAP7.
  Area denotes standard-cell area;
  rWL = routed wirelength;
  Pwr = total power;
  WNS / TNS = worst / total negative slack;
  DRC = DRC violations.\\
  ``w/o''~/~``w/'' denote results without / with the proposed initialization.
  Best result per (benchmark, metric) in \textbf{bold}.}
\label{tab:postroute}
\vspace{-6pt}
\begin{footnotesize}
\setlength{\tabcolsep}{4pt}
\begin{tabular}{l l r r r r r r}
\toprule
Benchmark & Configuration
  & Area ($\mu\mathrm{m}^2$) & rWL ($\mu\mathrm{m}$) & Pwr ($\mathrm{mW}$)
  & WNS ($\mathrm{ns}$) & TNS ($\mathrm{ns}$) & DRC \\
\midrule
\multirow{2}{*}{Ariane133}
  & DREAMPlace w/o init
    & 16392 & 982089 & 505.1
    & \textbf{$-$0.175} & \textbf{$-$276.2}
    & 157 \\
  & DREAMPlace w/ init
    & \textbf{16374} & \textbf{959926} & \textbf{504.0}
    & $-$0.263 & $-$472.7
    & \textbf{93} \\
\midrule
\multirow{2}{*}{BlackParrot}
  & DREAMPlace w/o init
    & 122960 & 7013052 & 1542.7
    & \textbf{$-$0.091} & \textbf{$-$277.0}
    & \textbf{4}\\
  & DREAMPlace w/ init
    & \textbf{122063} & \textbf{6926938} & \textbf{1541.3}
    & $-$0.103 & $-$366.1
    & \textbf{4}\\
\midrule
\multirow{2}{*}{\shortstack[l]{Mempool\\group}}
  & DREAMPlace w/o init
    & 336631 & 24499970 & 1363.0
    & \textbf{$-$0.127} & $-$678.6
    & 32576\\
  & DREAMPlace w/ init
    & \textbf{335528} & \textbf{24233201} & \textbf{1355.7}
    & $-$0.150 & \textbf{$-$440.2}
    & \textbf{32449}\\
\bottomrule
\end{tabular}
\end{footnotesize}
\end{table*}

\subsubsection{Timing Discussion}
While the proposed initialization consistently
improves routed wirelength, timing metrics
do not always improve in parallel.
On Ariane133, DREAMPlace w/ init achieves lower
wirelength but worse  WNS
($-$0.263\,ns vs.\ $-$0.175\,ns for w/o init).
This timing trade-off is expected: the initializer optimizes
a wirelength approximation and does not model timing paths directly.
Wirelength reduction benefits congestion
and routability but does not guarantee
timing improvement, which depends on
path-specific delays affected by cell
placement relative to clock domains.

\subsection{Summary of Findings}
\label{subsec:exp_summary}

The experimental results support three
conclusions.
(1)~The proposed Stiefel-constrained
initialization consistently reduces
pre-route wHPWL across three placement
flows: 3.5\% with DREAMPlace across all 20 benchmarks,
4.8\% with AutoDMP across its 19 converged baselines,
and up to 36.2\% with RePlAce on individual
designs.
This supports its generality as an initializer across 
downstream placement tools.
All converged runs show positive wHPWL improvement.
(2)~Pre-route improvements carry through the
full place-and-route flow: DREAMPlace w/ init reduces
post-route routed wirelength by 1.1--2.3\%
relative to w/o init on all three post-route benchmarks,
and also reduces standard-cell area and power on
all three.
DRC violations decrease on Ariane133 and Mempool
group and remain unchanged on BlackParrot.
Timing is mixed because the initializer optimizes
a wirelength approximation and does not model timing
paths.
(3)~The initialization can stabilize
hyperparameter-tuned analytical flows on
challenging designs.
On {\it adaptec3}, AutoDMP fails to converge with its
default initialization (N/C in
Table~\ref{tab:preroute_mms}), but converges
successfully when initialized by the proposed
method, producing a wHPWL comparable to
DREAMPlace.

\section{Conclusion}
\label{sec:conclusion}
This paper presented a placement initialization method for mixed-size
VLSI designs that formulates initialization as a QCQP 
with explicit first- and second-order constraints.
The key insight is that the quadratic equality constraints defining
area-weighted spreading already encode a Stiefel geometry.
Making this structure explicit via metric equalization transforms the
problem to a Stiefel manifold and enables feasibility-preserving
Riemannian optimization with Newton-CG second-order steps.
Combined with angle-sorted cycle sparsification and periodic
Laplacian updates, the method produces high-quality initial placements
that serve as a drop-in initializer for existing analytical placers.

Experiments across three state-of-the-art placers
confirm improvements on every converged benchmark.
On the TILOS benchmarks, the proposed initialization reduces
wHPWL by 13.8\% on average under RePlAce, with a 36.2\%
peak improvement on NVDLA\@.
DREAMPlace achieves a 3.5\% average improvement across all
20 benchmarks (14.0\% peak on newblue3).
AutoDMP achieves a 4.8\% average improvement across the
19 benchmarks with converged baselines (16.1\% peak on
newblue3) and additionally stabilizes the non-convergent
adaptec3 baseline.
All converged runs show positive improvement.
Post-route evaluation confirms that pre-route gains carry through:
routed wirelength decreases by 1.1--2.3\% on all three
post-route benchmarks, and standard-cell area and power decrease on all three.
DRC violations decrease on Ariane133 (40\% reduction) and Mempool group,
and remain unchanged on BlackParrot.
Timing is mixed because the initializer optimizes a wirelength approximation
and does not model timing paths.

Future directions focus on improving scalability and scope. For the implementation used in the reported experiments, initialization runtime ranges from 2.4 to 87.8 minutes on the 20-benchmark pre-route suite (Tables~\ref{tab:preroute_tilos} and~\ref{tab:preroute_mms}) and reaches about two hours on the Mempool group, motivating further optimization, including broader GPU acceleration beyond the current partial SSM support. Future work will incorporate congestion and timing terms into the initialization objective to target end-to-end PPA. Additional directions include adaptive subspace selection for faster convergence on large designs, post-route co-optimization with congestion- and timing-aware objectives, and extensions to three-dimensional placement.

\section*{Acknowledgments}
This work was supported in part by the National Science
Foundation under Grant CCF-2110419.



\bibliographystyle{IEEEtran}
\bibliography{bstctl,references}

@IEEEtranBSTCTL{IEEEexample:BSTcontrol,
  CTLuse_forced_etal       = "yes",
  CTLmax_names_forced_etal = "6",
  CTLnames_show_etal       = "1"
}

@inproceedings{ChenCCHLW23_ISPD,
  author    = {Pengwen Chen and Chung{-}Kuan Cheng and Albert Chern and Chester Holtz and Aoxi Li and Yucheng Wang},
  title     = {Placement Initialization via Sequential Subspace Optimization with Sphere Constraints},
  booktitle = {Proceedings of the 2023 International Symposium on Physical Design (ISPD)},
  year      = {2023},
  publisher = {ACM},
  url       = {https://doi.org/10.1145/3569052.3571877},
  doi       = {10.1145/3569052.3571877}
}

@article{ChengKKW19_RePlAce,
  author  = {Chung{-}Kuan Cheng and Andrew B. Kahng and Ilgweon Kang and Lutong Wang},
  title   = {{RePlAce}: Advancing Solution Quality and Routability Validation in Global Placement},
  journal = {IEEE Transactions on Computer-Aided Design of Integrated Circuits and Systems},
  volume  = {38},
  number  = {9},
  pages   = {1717--1730},
  year    = {2019},
  url     = {https://doi.org/10.1109/TCAD.2018.2859220},
  doi     = {10.1109/TCAD.2018.2859220}
}

@article{LuCCSHTC15_ePlace,
  author  = {Jingwei Lu and Pengwen Chen and Chin{-}Chih Chang and Lu Sha and Dennis Jen{-}Hsin Huang and Chin{-}Chi Teng and Chung{-}Kuan Cheng},
  title   = {{ePlace}: Electrostatics-Based Placement Using Fast Fourier Transform and {Nesterov}'s Method},
  journal = {ACM Transactions on Design Automation of Electronic Systems},
  volume  = {20},
  number  = {2},
  pages   = {17:1--17:34},
  year    = {2015},
  url     = {https://doi.org/10.1145/2699873},
  doi     = {10.1145/2699873}
}

@inproceedings{LinDLRKP19_DREAMPlace,
  author    = {Yibo Lin and Shounak Dhar and Wuxi Li and Haoxing Ren and Brucek Khailany and David Z. Pan},
  title     = {{DREAMPlace}: Deep Learning Toolkit-Enabled {GPU} Acceleration for Modern {VLSI} Placement},
  booktitle = {Proceedings of the 56th Annual Design Automation Conference (DAC)},
  year      = {2019},
  publisher = {ACM},
  url       = {https://doi.org/10.1145/3316781.3317803},
  doi       = {10.1145/3316781.3317803}
}

@book{AbsilMahonySepulchre08,
  author    = {Pierre{-}Antoine Absil and Robert E. Mahony and Rodolphe Sepulchre},
  title     = {Optimization Algorithms on Matrix Manifolds},
  publisher = {Princeton University Press},
  year      = {2008},
  isbn      = {978-0-691-13298-3}
}

@article{EdelmanAriasSmith98,
  author  = {Alan Edelman and Tom{\'a}s A. Arias and Steven T. Smith},
  title   = {The geometry of algorithms with orthogonality constraints},
  journal = {SIAM Journal on Matrix Analysis and Applications},
  volume  = {20},
  number  = {2},
  pages   = {303--353},
  year    = {1998},
  url     = {https://doi.org/10.1137/S0895479895290954},
  doi     = {10.1137/S0895479895290954}
}

@misc{ChenCH24_StiefelSSM_arXiv,
  author       = {Pengwen Chen and Chung{-}Kuan Cheng and Chester Holtz},
  title        = {Sequential Subspace Methods on {Stiefel} Manifold Optimization},
  howpublished = {arXiv preprint arXiv:2404.13301},
  year         = {2024},
  eprint       = {2404.13301},
  archivePrefix= {arXiv},
  primaryClass = {math.OC}
}

@article{MarkovHuKim15_ProcIEEE_PlacementSurvey,
  author  = {Igor L. Markov and Jin Hu and Myung-Chul Kim},
  title   = {Progress and Challenges in {VLSI} Placement Research},
  journal = {Proceedings of the IEEE},
  volume  = {103},
  number  = {11},
  pages   = {1985--2003},
  year    = {2015},
  doi     = {10.1109/JPROC.2015.2478963}
}

@article{LuZhuangChenEtAl15_TCAD_ePlaceMS,
  author  = {Jingwei Lu and Hao Zhuang and Pengwen Chen and Hongliang Chang and Chin-Chih Chang and Yiu-Chung Wong and Lu Sha and Dennis J.-H. Huang and Yufeng Luo and Chin-Chi Teng and Chung-Kuan Cheng},
  title   = {{ePlace-MS}: Electrostatics-Based Placement for Mixed-Size Circuits},
  journal = {IEEE Transactions on Computer-Aided Design of Integrated Circuits and Systems},
  volume  = {34},
  number  = {5},
  pages   = {685--698},
  year    = {2015},
  doi     = {10.1109/TCAD.2015.2391263}
}

@inproceedings{YanViswanathanChu09_DAC_MixedSizeComplexities,
  author={Yan, Jackey Z. and Viswanathan, Natarajan and Chu, Chris},
  booktitle={2009 46th ACM/IEEE Design Automation Conference}, 
  title={Handling complexities in modern large-scale mixed-size placement}, 
  year={2009},
  volume={},
  number={},
  pages={436-441},
  doi={10.1145/1629911.1630028}}

@article{Mirhoseini21_Nature_GraphPlacement,
  author  = {Azalia Mirhoseini and Anna Goldie and Mustafa Yazgan and Joe Wenjie Jiang and Ebrahim M. Songhori and Shen Wang and Young-Joon Lee and Eric Johnson and Omkar Pathak and Azade Nazi and Jiwoo Pak and Andy Tong and Kavya Srinivasa and William Hang and Emre Tuncer and Quoc V. Le and James Laudon and Richard Ho and Roger Carpenter and Jeff Dean},
  title   = {A graph placement methodology for fast chip design},
  journal = {Nature},
  volume  = {594},
  number  = {7862},
  pages   = {207--212},
  year    = {2021},
  doi     = {10.1038/S41586-021-03544-W},
  url     = {https://doi.org/10.1038/s41586-021-03544-w}
}

@article{Liu22_TODAES_GraphPlanner,
  author  = {Yiting Liu and Ziyi Ju and Zhengming Li and Mingzhi Dong and Hai Zhou and Jia Wang and Fan Yang and Xuan Zeng and Li Shang},
  title   = {{GraphPlanner}: Floorplanning with Graph Neural Network},
  journal = {ACM Transactions on Design Automation of Electronic Systems},
  volume  = {28},
  number  = {2},
  pages   = {21:1--21:24},
  year    = {2023},
  doi     = {10.1145/3555804},
  url     = {https://doi.org/10.1145/3555804}
}

@inproceedings{LiuZhouWangEtAl24_GiFt,
  author={Liu, Yiting and Zhou, Hai and Wang, Jia and Yang, Fan and Zeng, Xuan and Shang, Li},
  booktitle={2024 ACM/IEEE International Conference On Computer Aided Design (ICCAD)}, 
  title={The Power of Graph Signal Processing for Chip Placement Acceleration}, 
  year={2024},
  volume={},
  number={},
  pages={1-8},
  doi={10.1145/3676536.3676669}}

@inproceedings{AutoDMP23_ISPD,
  author    = {Anthony Agnesina and Puranjay Rajvanshi and Tian Yang and Geraldo Pradipta and Austin Jiao and Ben Keller and Brucek Khailany and Haoxing Ren},
  title     = {{AutoDMP}: Automated {DREAMPlace}-based Macro Placement},
  booktitle = {Proceedings of the 2023 International Symposium on Physical Design (ISPD)},
  year      = {2023},
  pages     = {149--157},
  publisher = {ACM},
  doi       = {10.1145/3569052.3578923},
  url       = {https://doi.org/10.1145/3569052.3578923}
}

@inproceedings{LiaoEtAl_StrongerMixedSize,
  author={Chen, Yifan and Wen, Zaiwen and Liang, Yun and Lin, Yibo},
  booktitle={2023 IEEE/ACM International Conference on Computer Aided Design (ICCAD)}, 
  title={Stronger Mixed-Size Placement Backbone Considering Second-Order Information}, 
  year={2023},
  volume={},
  number={},
  pages={1-9},
  doi={10.1109/ICCAD57390.2023.10323700}}

@article{KahngEtAl_UpdatedAssessment,
  author={Cheng, Chung-Kuan and Kahng, Andrew B. and Kundu, Sayak and Wang, Yucheng and Wang, Zhiang},
  journal={IEEE Transactions on Computer-Aided Design of Integrated Circuits and Systems}, 
  title={An Updated Assessment of Reinforcement Learning for Macro Placement}, 
  year={2025},
  volume={},
  number={},
  pages={1-1},
  doi={10.1109/TCAD.2025.3644293}}

@misc{TILOS_MacroPlacement,
  author       = {{TILOS AI Institute}},
  title        = {{MacroPlacement}},
  howpublished = {\url{https://github.com/TILOS-AI-Institute/MacroPlacement}},
  year         = {2023},
  note         = {Accessed: 2026-04-30}
}

@article{ClarkEtAl16_ASAP7,
  author    = {Lawrence T. Clark and Vinay Vashishtha and Lucian Shifren and Aditya Gujja and Saurabh Sinha and Brian Cline and Chandarasekaran Ramamurthy and Greg Yeric},
  title     = {{ASAP7}: A 7-nm finFET predictive process design kit},
  journal   = {Microelectronics Journal},
  volume    = {53},
  pages     = {105--115},
  year      = {2016},
  publisher = {Elsevier},
  doi       = {10.1016/j.mejo.2016.04.006}
}

@misc{Nangate45_OCL,
  author       = {{NanGate Inc.}},
  title        = {{NanGate FreePDK45 Open Cell Library}},
  howpublished = {\url{https://si2.org/open-cell-library/}},
  year         = {2008},
  note         = {Accessed: 2026-05-04}
}

\vspace{-0.8in}
\begin{IEEEbiography}[{\raisebox{0.3in}{\includegraphics[height=1.0in,
clip,keepaspectratio]
{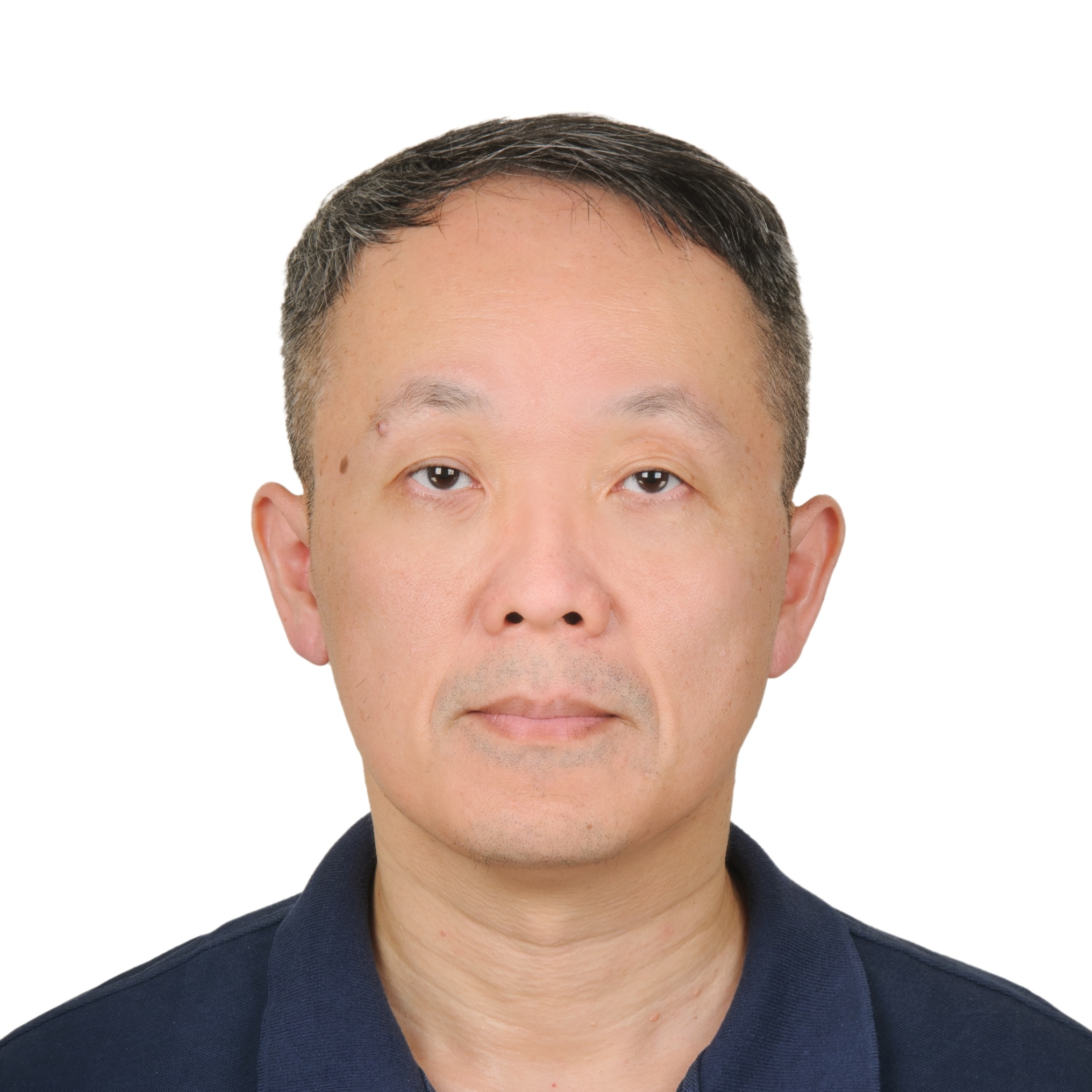}}}]{Pengwen Chen} is currently a  Professor in the Mathematics Department of  National Tsing Hua University, Hsinchu, Taiwan. 
His research interests include numerical analysis, optimization, and mathematical imaging science. 
He received his Ph.D. in Mathematics from the University of Florida.
\end{IEEEbiography}

\vspace{-0.9in}

\begin{IEEEbiography}[{\raisebox{0.3in}{\includegraphics[height=1.0in,clip,keepaspectratio]
{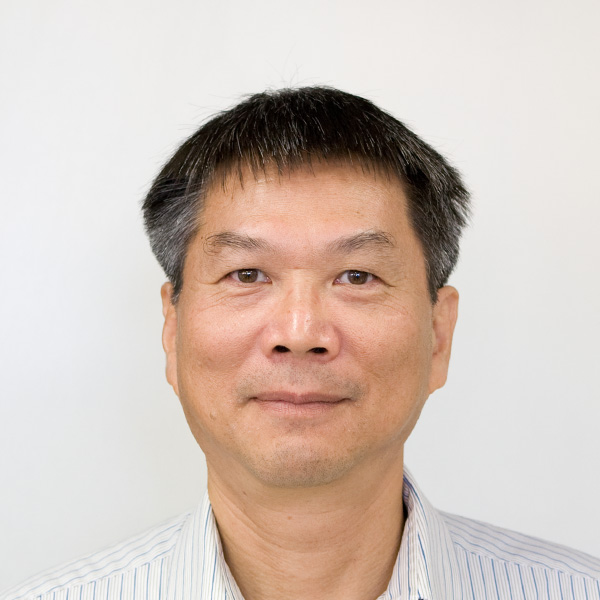}}}]{Chung-Kuan Cheng}
is a Distinguished Professor in the 
CSE Department and an Adjunct Professor in the 
ECE Department of the University of California, San Diego.
His research interests include computer-aided design for machine learning and design automation in microelectronic circuits. He received his Ph.D. degree in Electrical Engineering and Computer Sciences from the University of
California, Berkeley.
\end{IEEEbiography}

\vspace{-0.7in}

\begin{IEEEbiography}[{\raisebox{0.3in}{\includegraphics[
height=1.0in,
clip
]
{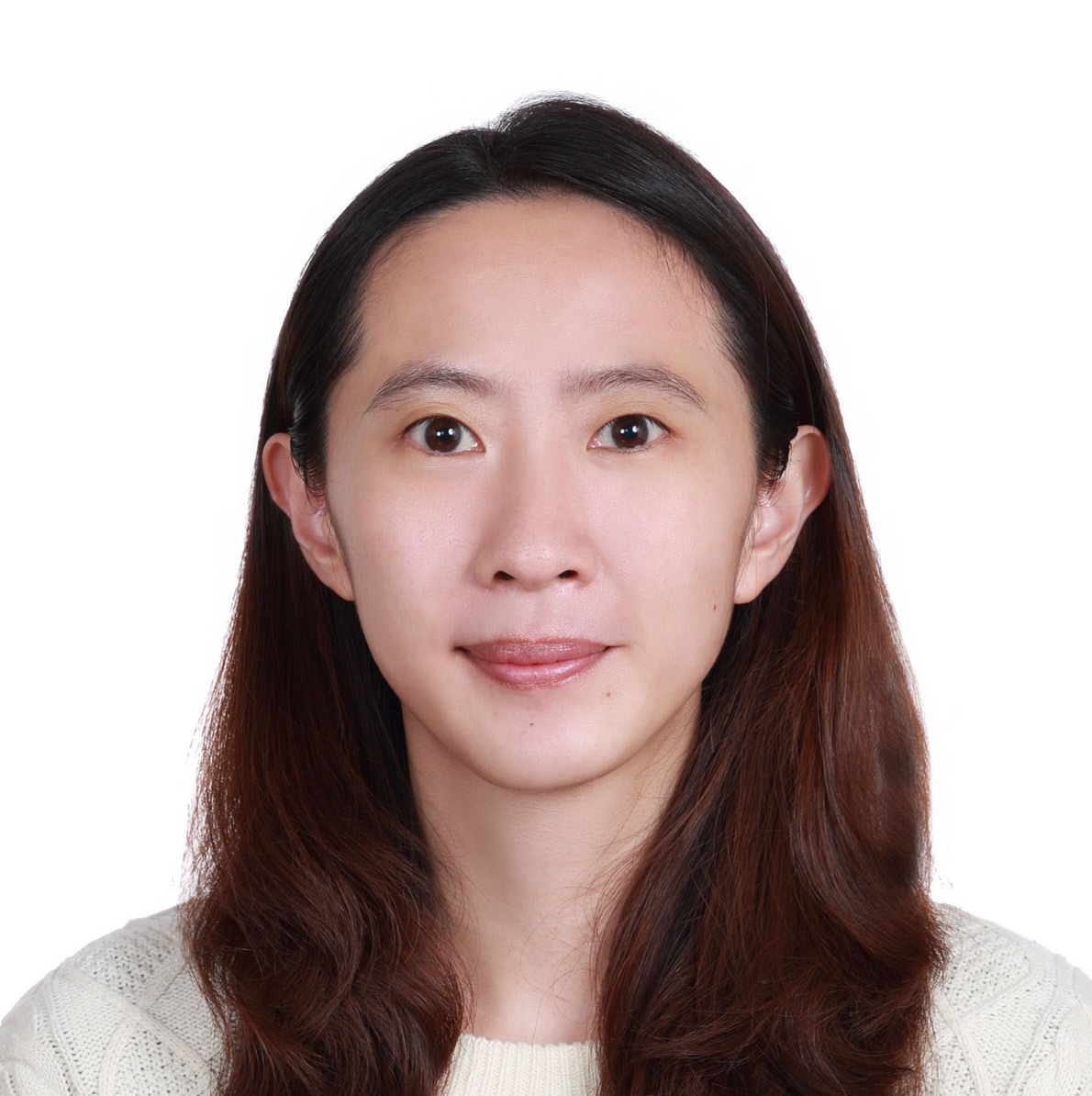}}}]{Shao-Yun Fang} received a B.S. degree in electrical engineering from the National Taiwan University (NTU), Taipei, Taiwan, in 2008 and a Ph.D. degree from the Graduate Institute of Electronics Engineering, NTU, in 2013. She is currently a Professor of the Department of Electrical Engineering, National Taiwan University of Science and Technology, Taipei, Taiwan. Her research interests focus on physical design and design for manufacturability for
\end{IEEEbiography}
\vspace{-0.14in}
{\footnotesize
\noindent integrated circuits. She is serving as an Associate Editor of IEEE Transactions on Computer-Aided Design of Integrated Circuits and Systems and ACM Transactions on Design Automation of Electronic Systems.
\par}

\vspace{-0.4in}

\begin{IEEEbiography}[
{\raisebox{0.3in}{\includegraphics[height=1.0in,clip,keepaspectratio]
{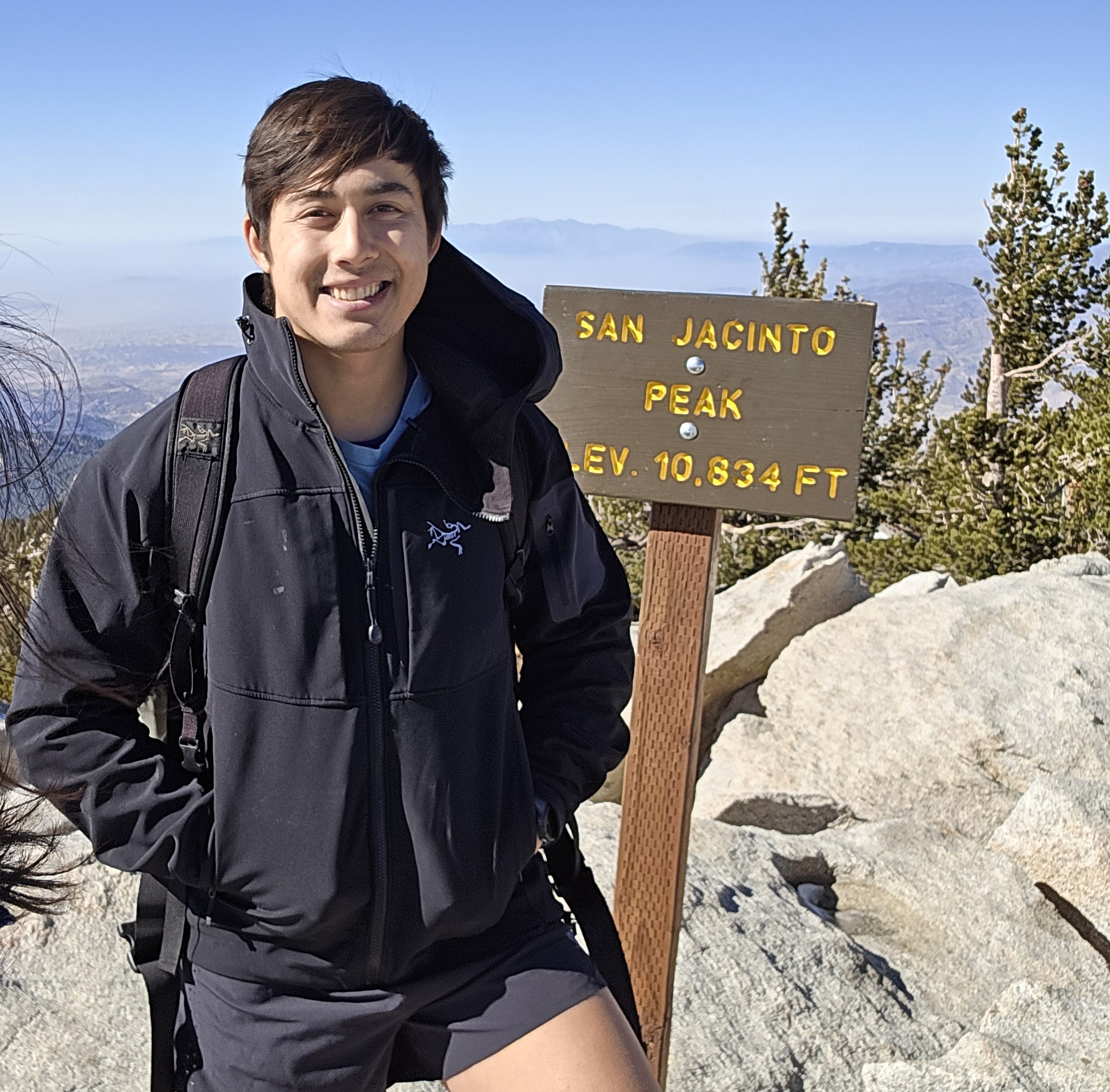}}}
]{Chester Holtz} is a Postdoctoral Scholar at the Halıcıoğlu Data Science Institute. He earned his Ph.D. in Computer Science from the University of California, San Diego where he was advised by Professor Chung-Kuan Cheng. His research interests include machine learning and optimization, in particular on graph-structured data. 
\end{IEEEbiography}

\vspace{-0.7in}

\begin{IEEEbiography}[{\raisebox{0.3in}{\includegraphics[height=1.0in,clip,keepaspectratio]
{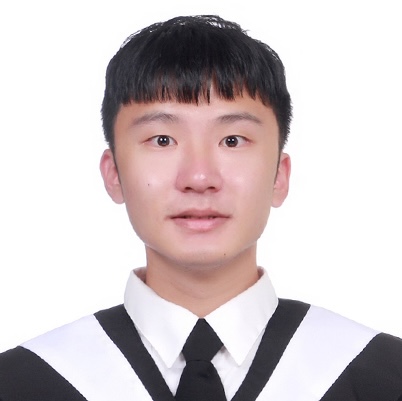}}}]{Sheng-Tan Huang} received a B.S. degree in Electrical Engineering from National Taiwan University of Science and Technology (NTUST), Taipei, Taiwan, in 2022. He is currently pursuing a Ph.D. degree in Electrical Engineering at National Taiwan University of Science and Technology, Taipei, Taiwan. His current research interests focus on quantum layout synthesis and large-scale placement.
\end{IEEEbiography}

\vspace{-0.6in}

\begin{IEEEbiography}[{\raisebox{0.3in}{\includegraphics[height=1.0in,clip,keepaspectratio]
{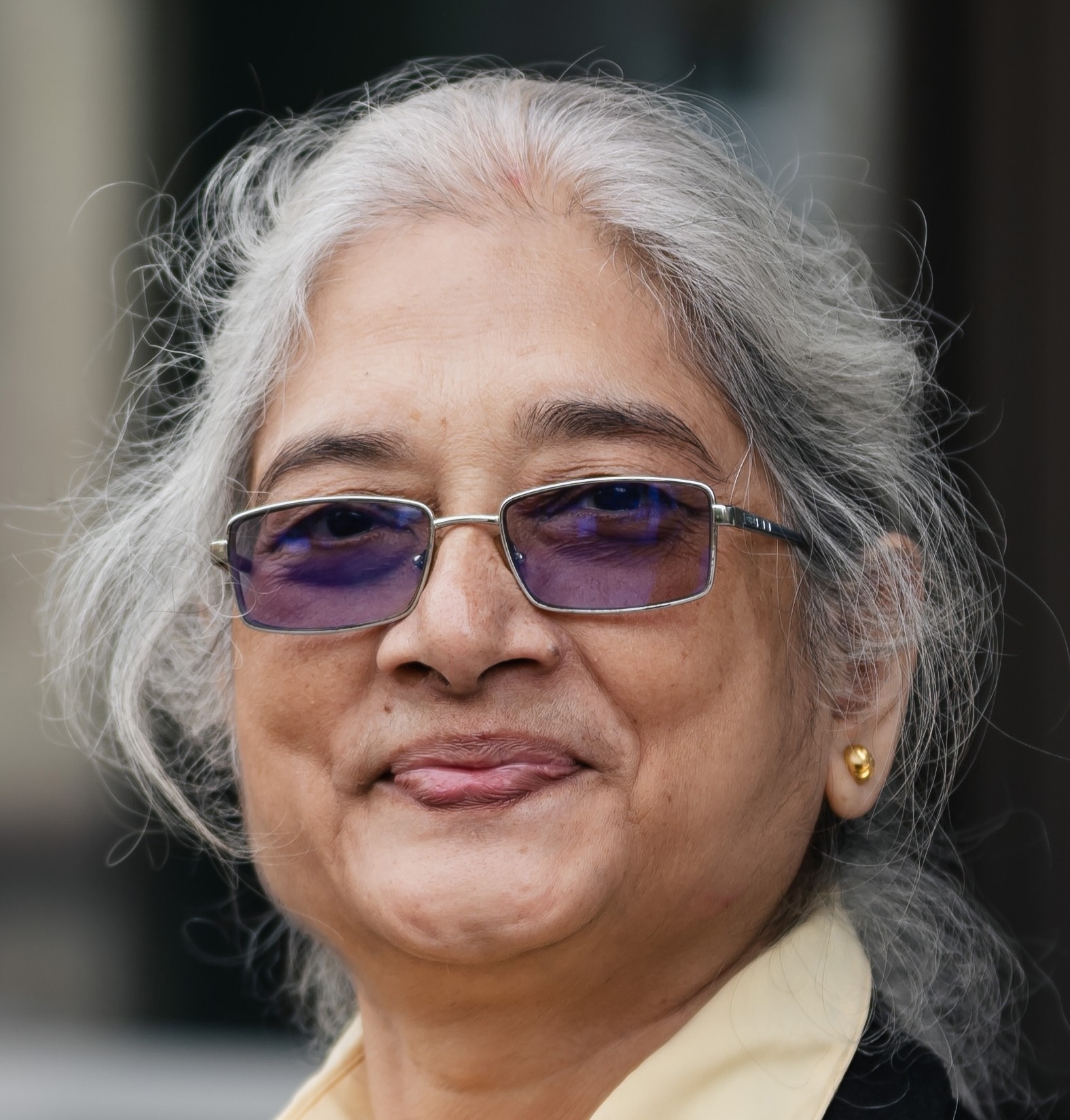}}}]{Susmita Sur-Kolay}
is a Fulbright Academic Excellence Fellow in the 
CSE Department of the University of California, San Diego and has been a Professor at Indian Statistical Institute.
Her research interests include design automation for microelectronic and quantum circuits, hardware security and in-memory neuromorphic computation. She received her Ph.D. degree in Computer Science and Engineering from Jadavpur University, and from 1980 to 
\end{IEEEbiography}
\vspace{-0.14in}
{\footnotesize
\noindent 1985, she was at Massachusetts Institute of Technology. She is a Fellow of Indian National Science Academy and Indian National Academy of Engineering, a recipient of the President of India Gold Medal and Distinguished
Alumnus Award of IIT Kharagpur, and Women in Technology Leadership Award of the VLSI Society of India.
\par}

\vspace{-0.4in}

\begin{IEEEbiography}[{\raisebox{0.3in}{\includegraphics[height=1.0in,clip,keepaspectratio]
{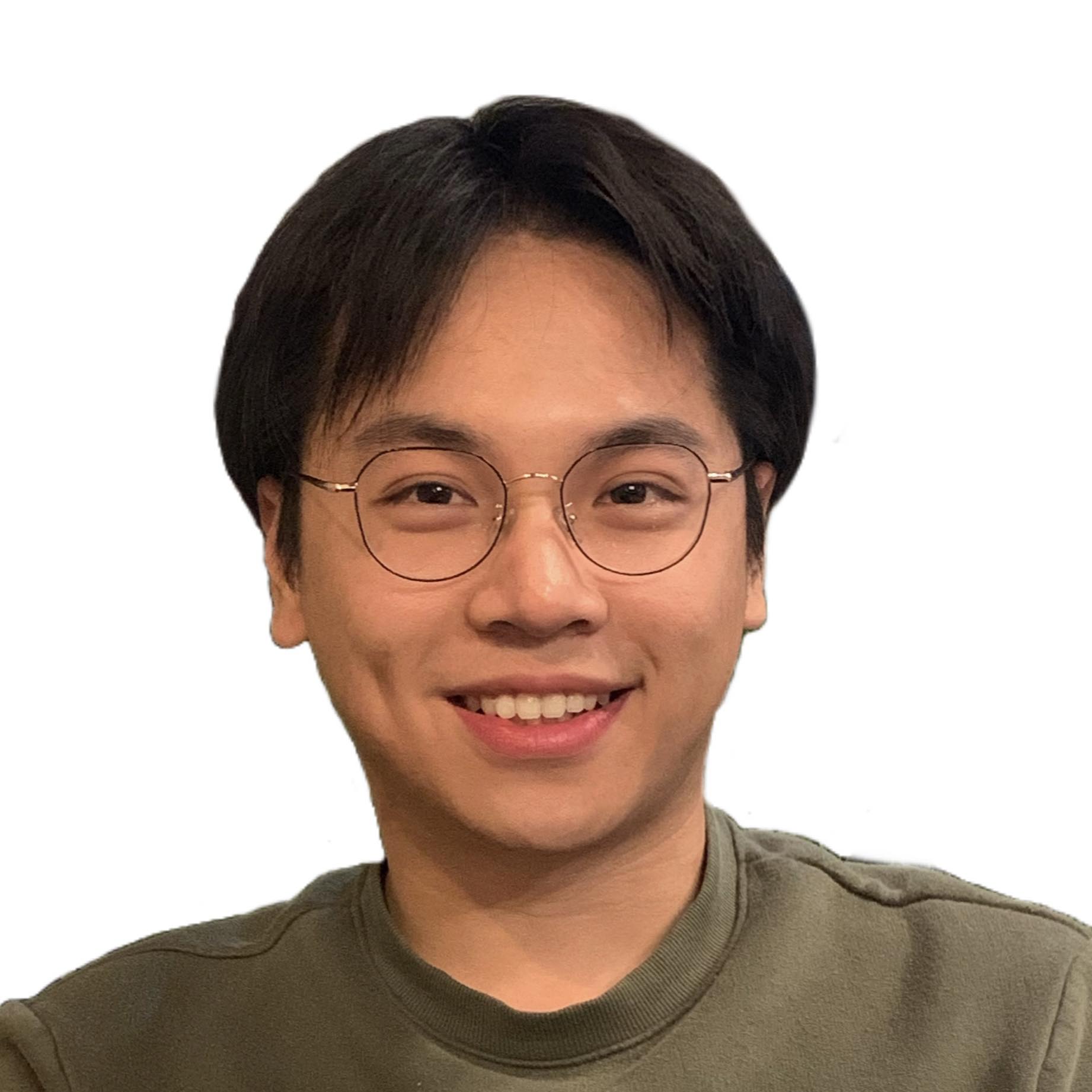}}}]{Yucheng Wang} 
is a Ph.D. student in the Department of Computer Science and Engineering, University of California, San Diego. He received his bachelor's degree in Computer Science from Purdue University, West Lafayette, in 2021. His research interests include standard-cell layout automation, design technology co-optimization strategies, and graph visualization.
\end{IEEEbiography}

\vspace{-0.6in}

\begin{IEEEbiography}[
{\raisebox{0.3in}{\includegraphics[height=1.0in,clip,keepaspectratio]
{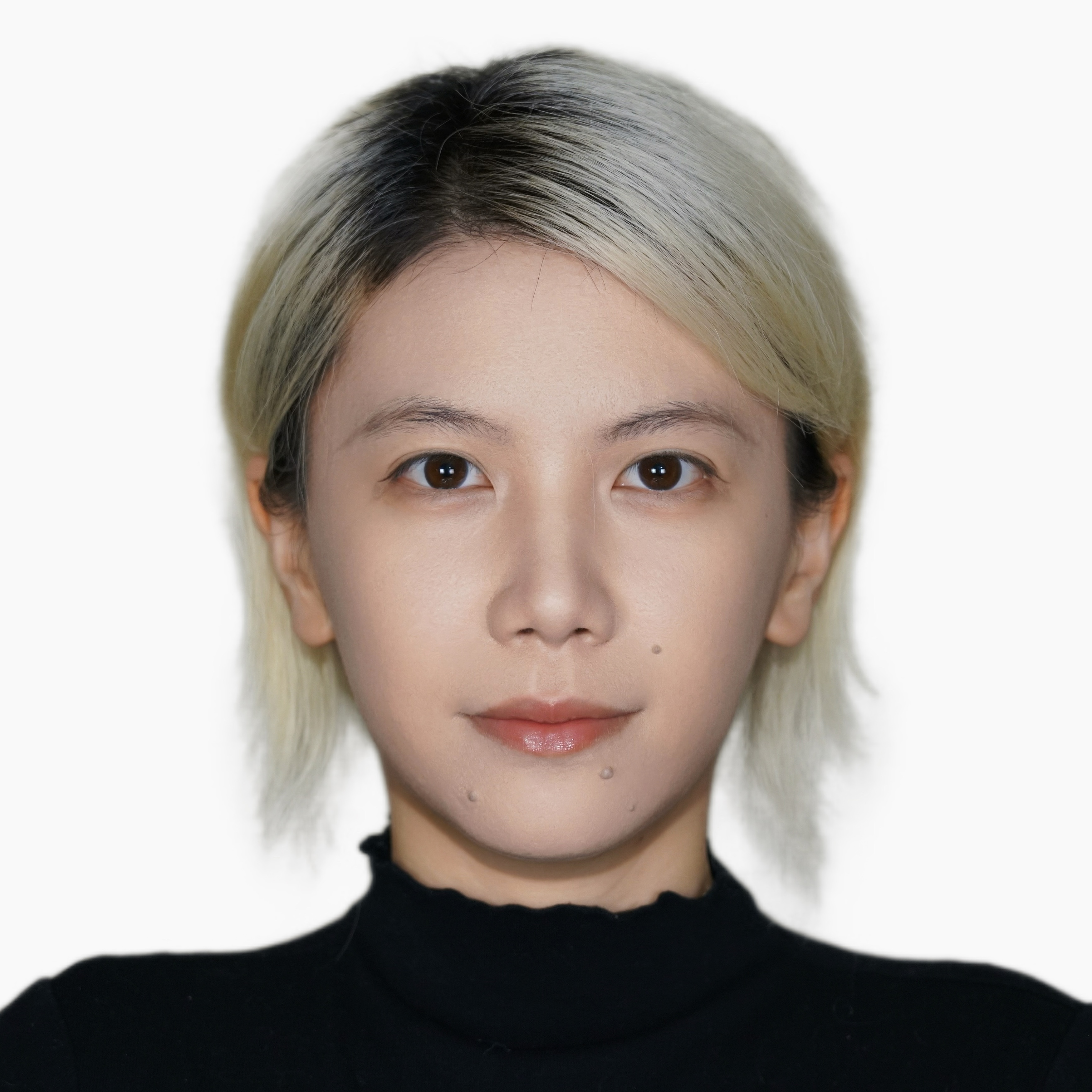}}}
]{Ying Yuan} is a Ph.D. student in the Department of Computer Science and Engineering, University of California, San Diego. She received her bachelor's degree in Computer Science from Purdue University, West Lafayette, in 2021, and her master's degree in Computer Science and Engineering from the University of Michigan, Ann Arbor, in 2023. Her research interests include 2D and 3D placement and routing. Her work integrates machine learning and optimization.
\end{IEEEbiography}

\end{document}